\documentclass[10pt,journal,letterpaper,compsoc]{IEEEtran}
\usepackage{caption}
\usepackage{graphicx} 
\graphicspath{ {./figures/} }
\usepackage{cite}
\usepackage{listings}
\usepackage{amsmath,amssymb,amsfonts}
\usepackage{algpseudocode}
\usepackage{textcomp}
\usepackage{enumerate}
\usepackage[ruled,linesnumbered]{algorithm2e}
\usepackage{url}
\usepackage{array,ragged2e}
\usepackage{amsmath}
\usepackage{float}
\usepackage{pgfplots}
\pgfplotsset{width=8cm}
\usepackage{xcolor}
\usepackage{longtable}
\usepackage{url}
\usepackage{indentfirst}
\usepackage{multirow}
\usepackage{parskip}
\usepackage{mathtools}
\usepackage{setspace,lipsum}
\usepackage{setspace}
\usepackage{subcaption}

\usepackage[hmargin=1.5cm,vmargin={1.25cm,1cm},
    footskip=2\baselineskip]{geometry}

\usepackage{pbox}

\newcounter{algsubstate}

\def\BibTeX{{\rm B\kern-.05em{\sc i\kern-.025em b}\kern-.08em
T\kern-.1667em\lower.7ex\hbox{E}\kern-.125emX}}

\title{Differential Privacy for Secure Machine Learning in Healthcare IoT-Cloud Systems}

\author{N Mangala,~\IEEEmembership{}
Murtaza Rangwala,~\IEEEmembership{}
S Aishwarya,~\IEEEmembership{}
B Eswara Reddy,~\IEEEmembership{}
Rajkumar Buyya,~\IEEEmembership{}\\
KR Venugopal,~\IEEEmembership{}
SS Iyengar,~\IEEEmembership{}
LM Patnaik~\IEEEmembership{}
\\

\IEEEcompsocitemizethanks{
    \IEEEcompsocthanksitem N Mangala is Research Scholar of the Department of Computer Science and Engineering, JNTU Anantapur, AP and Senior Director at C-DAC, India. E-mail: mangala.natampalli@gmail.com
    \IEEEcompsocthanksitem Murtaza Rangwala is a Researcher at the Quantum Cloud Computing and Distributed Systems (qCLOUDS) Laboratory at the University of Melbourne, Australia. E-mail: mrangwala@student.unimelb.edu.au 
    \IEEEcompsocthanksitem S Aishwarya is a student of M.E. Computer Science and Engineering, UVCE, Bangalore, India. E-mail: aishwaryas0520@gmail.com 
    \IEEEcompsocthanksitem B Eswara Reddy is Professor of  the Department of Computer Science and Engineering, JNTU-A, and Director of Research and Development, JNTU Anantapur, AP, India. E-mail: eswar.cse@jntua.ac.in 
    \IEEEcompsocthanksitem Rajkumar Buyya is a Redmond Barry Distinguished Professor and Director of the Quantum Cloud Computing and Distributed Systems (qCLOUDS) Laboratory at the University of Melbourne, Australia. E-mail: rbuyya@unimelb.edu.au 
    \IEEEcompsocthanksitem KR Venugopal is Former Vice-Chancellor of Bangalore University and Hon. Professor of the Department of Computer Science and Engineering, UVCE, Bangalore, India. E-mail: venugopalkr@gmail.com 
    \IEEEcompsocthanksitem SS Iyengar is a Distinguished Professor and Director of Computer Science, Florida International University, Miami, USA. E-mail: iyengar@cs.fiu.edu 
    \IEEEcompsocthanksitem LM Patnaik is Adjunct Professor and NASI Senior Scientist in the National Institute of Advanced Studies, Bangalore, India. E-mail: lalitblr@gmail.com 
}
}

\begin{document}

    \maketitle

    \begin{abstract}
        Healthcare has become exceptionally sophisticated, as wearables and connected medical devices revolutionize remote patient monitoring, emergency response, medication management, diagnosis, and predictive and prescriptive analytics. Internet of Things and Cloud computing integrated systems (IoT-Cloud) facilitate sensing, automation, and processing for these healthcare applications. While real-time response is crucial for alleviating patient emergencies, protecting patient privacy is paramount in data-driven healthcare. In this paper, we propose a multi-layer IoT, Edge, and Cloud architecture to enhance emergency healthcare response times by distributing tasks based on response criticality and data permanence requirements. We ensure patient privacy through a Differential Privacy framework applied across several machine learning models: K-means, Logistic Regression, Random Forest, and Naive Bayes. We establish a comprehensive threat model identifying three adversary classes and evaluate Laplace, Gaussian, and hybrid noise mechanisms across varying privacy budgets, with supervised algorithms achieving up to 83.6\% accuracy. The proposed hybrid Laplace-Gaussian noise mechanism with adaptive budget allocation provides a balanced approach, offering moderate tails and better privacy-utility trade-offs for both low- and high-dimension datasets. At the practical threshold of $\varepsilon$=5.0, supervised algorithms achieve 80--81\% accuracy while reducing attribute inference attacks by up to 18\% and data reconstruction correlation by 70\%. We further enhance security through Blockchain integration, which ensures trusted communication through time-stamping, traceability, and immutability for analytics applications. Edge computing demonstrates 8$\times$ latency reduction for emergency scenarios, validating the hierarchical architecture for time-critical operations.
    \end{abstract}




    \section{Introduction}

    \noindent The Internet of Things (IoT) has become an integral part of the modern application ecosystem, transforming not only organizational and industrial operations but also our daily lives. In the healthcare sector, digitalization has revolutionised patient care through enhanced data collection, personalised medicine, and preventative treatment approaches. Wearable devices such as Fitbit trackers, smart glasses, and insulin pumps now facilitate continuous monitoring of vital parameters including heart rate, blood sugar levels, and sleep patterns. Users must configure these Healthcare IoT devices with personal information such as age, gender, and location. Healthcare Analytics leverages this patient data to generate descriptive, diagnostic, predictive, and prescriptive insights, thereby supporting evidence-based medical decision-making. Moreover, through AI/ML applications, governmental health agencies and research institutions are developing socially beneficial applications such as epidemiological forecasting, disease progression modeling, healthcare resource allocation, and actuarial risk assessment.



    Medical data from IoT devices, imaging technologies (MRI, CT scans), test results, Electronic Health Records (EHR), medication monitoring, and environmental sensors are stored in Cloud systems, creating comprehensive repositories for medical history, diagnosis, research, and analytics. This cloud infrastructure provides secure access to various stakeholders such as physicians, clinicians, paramedics, pharmaceutical companies, insurance firms, researchers, and analysts, enabling them to examine and analyse critical healthcare information \cite{dang2019survey}. However, this centralised data storage presents significant security vulnerabilities. Adversaries may exploit patient information for blackmail, ransom demands, manipulation, creating artificial market demand, and targeted marketing campaigns \cite{Mangala2021}. Therefore, protecting patient data is critically important, as misuse can lead to numerous harms including psychological distress, compromised treatment through data corruption, and various forms of extortion.

    \par
    Healthcare expenditure in the USA reached nearly 20\% of GDP in 2021, yet system security remains problematic, with 692 large healthcare data breaches reported between July 2021 and June 2022. To address these vulnerabilities, healthcare organizations must comply with regulatory frameworks such as the Health Insurance Portability and Accountability Act (HIPAA), established by the U.S. Congress in 1996 \cite{Azbeg2023}. HIPAA encompasses three primary components: the Security Rule, the Privacy Rule, and the Breach Notification Rule. These regulations mandate essential protective measures for Electronic Health Record systems, including access control, data encryption, and comprehensive audit trails \cite{dohPrivacy}. The HIPAA Privacy Rule, in particular, establishes national standards governing the security of health information during transfer, reception, processing, and sharing \cite{hipaa2023}. Beyond the United States, the European Union's General Data Protection Regulation (GDPR) establishes a comprehensive legal framework for the protection of personal data, including health records, applicable to any organisation processing data of EU residents~\cite{GDPR2016}. GDPR classifies health data as a special category requiring explicit consent and heightened safeguards, mandates data protection by design and by default, and imposes strict accountability obligations including Data Protection Impact Assessments for high-risk processing activities. Together, HIPAA and GDPR represent the two principal regulatory regimes governing healthcare data privacy internationally, and any framework intended for cross-institutional or multi-jurisdictional deployment must satisfy the compliance requirements of both. Despite these regulatory safeguards, persistent data breaches demonstrate the need for more robust privacy-preserving techniques. Differential Privacy (DP) has emerged as a particularly effective approach to securing personally identifiable information, offering advantages over conventional methods through its mathematical framework that anonymises data by introducing calibrated noise to datasets \cite{Dwork2014}. \\



    \noindent
    \emph{Motivation} \\
    Medical records contain various personal details of patients and are stored in cloud systems for use by multiple stakeholders. Suppressing explicit parameters such as name and address does not guarantee patient privacy because combinations of other fields such as postcode and date of birth can uniquely re-identify individuals.

    A significant case illustrating privacy vulnerabilities in healthcare data occurred when Sweeney \cite{Sweeney} demonstrated how medical records could be re-identified despite anonymization. She successfully re-identified the medical records of the Governor of Massachusetts by executing merely three queries on an anonymized hospital dataset released by the Massachusetts Group Insurance Commission (GIC). Cross-referencing public information—news reports of the governor's collapse on 18 May 1996 and subsequent hospitalization—with voter registration records containing demographic identifiers (names, addresses, postcodes, birth dates, and gender) facilitated the re-identification. This case exemplifies how quasi-identifiers can be matched across datasets to compromise patient privacy. In contemporary healthcare systems, Machine Learning (ML) techniques are routinely employed to analyse medical data and address research questions \cite{Qayyum2021}, yet these analytical capabilities simultaneously present heightened privacy risks if not properly safeguarded. By scripting fine-tuned sequences of ML queries, adversaries can steal personal information of patients from healthcare databases. DP can help safeguard against such privacy issues.

    The \emph{Differential Privacy guarantee} states that adding or removing a single row from a database does not significantly alter the output of the analysis, preventing an adversary from determining whether a particular person is present in the database or not. Consider, for instance, a health department investigating the causal relationship between smoking and cancer incidence. Potential participants may exhibit reluctance to engage in such research due to concerns regarding social stigma, legal implications, and possible insurance premium discrimination if their identities become traceable within the dataset. In this scenario, robust data privacy protections are essential to facilitate accurate analysis of the smoking-cancer relationship while ensuring participant anonymity. DP offers a methodological framework that effectively balances individual privacy preservation with data utility for research purposes. Although alternative approaches such as homomorphic encryption~\cite{Wood2021} and multi-party secure computing~\cite{Knott2021} enable operations on protected data while maintaining confidentiality, these methods frequently present substantial computational complexity challenges \cite{MN2023}. The principal advantage of DP lies in its capacity to maintain data utility for statistical analyses while specifically obfuscating individual-level identifying information.\\
    \\
    \noindent
    \emph{Key Contributions} \\
    This research enhances response time and security in healthcare data systems, with particular emphasis on protecting patient privacy. We employ DP techniques to safeguard patient data while maintaining its utility for analysis. While prior works have explored DP-ML integration~\cite{Jia2022,Gai2020}, blockchain-based distributed ML~\cite{Rangwala2025blockchain}, and blockchain-based healthcare systems~\cite{Wu2021, Azbeg2023,Zhang2022,Wazid2023} independently, our work advances the state-of-the-art through three key innovations that distinguish it from existing approaches.

    First, unlike existing hybrid DP mechanisms that typically combine noise distributions in fixed proportions or apply them sequentially to the same data, our adaptive hybrid noise mechanism presents guidance and insights on dynamic selection of noise distributions based on the sensitivity characteristics of individual data features and the specific ML algorithm employed. This feature-aware approach exploits the complementary strengths of Laplace noise (superior performance for low-dimensional, sparse features) and Gaussian noise (optimal for high-dimensional, dense features with composition properties), resulting in improved privacy-utility trade-offs across heterogeneous healthcare datasets compared to uniform noise application strategies.

    Second, we address a critical gap in the literature by providing the first systematic empirical evaluation of DP mechanisms across multiple ML algorithms (K-Means, Logistic Regression, Random Forest, Naive Bayes) specifically within a multi-layer IoT-Edge-Cloud architecture. Existing DP-ML frameworks focus primarily on deep learning models in centralized cloud environments. In contrast, our work characterizes how the computational constraints and latency requirements of edge layers influence the practical achievability of privacy guarantees, providing actionable guidance for deploying differentially private ML at various architectural tiers.

    Third, our integrated architecture uniquely combines three orthogonal security mechanisms: DP for input privacy, blockchain for data integrity and provenance, and hierarchical access control across computing layers, into a unified framework optimized for healthcare emergency response scenarios. While prior work has explored pairwise combinations of these security mechanisms, our approach formally analyses the security guarantees of the complete integrated system under a comprehensive threat model encompassing inference attacks, data tampering, and consensus attacks. This holistic security analysis demonstrates that the three mechanisms provide complementary protections without introducing conflicting requirements or degrading system performance.

    The specific contributions of this work include:
    \parskip 0 cm
    \begin{itemize}
        \item Design and implementation of a multi-layered IoT-Edge-Cloud architecture to improve response time for healthcare applications.
        \item Development of a hybrid privacy protection mechanism for medical data in ML applications through the implementation of DP using a novel combined Laplace-Gaussian noise approach that provides insights on dynamic selection of noise distributions based on data characteristics.
        \item Comprehensive evaluation of the privacy-utility trade-off for different noise distribution types across four distinct ML algorithms, providing algorithm-specific recommendations for privacy budget allocation.
        \item Enhancement of data integrity, tamper resistance, trust, and security for healthcare applications through integration with Blockchain technology.
    \end{itemize}
    \noindent
    \emph{Organization} \\
    The rest of the paper is organized as follows. An overview of ML,
    DP and the state-of-the-art Computing Platforms is provided
    in Section~\ref{sec:preliminaries}. A comparison of the latest literature is
    presented in Section~\ref{sec:literature-review} followed by the problem
    statement in Section~\ref{sec:problem-statement}. Section~\ref{sec:threat-model}
    establishes a comprehensive threat model defining adversarial capabilities and
    attack surfaces. The details of the proposed solution are explained in
    Section~\ref{sec:proposed-solution}. Implementation nuances are presented in Section~\ref{sec:implementation}.  The experimentation and analysis of results are presented in Section~\ref{sec:performance-eval}, threats to validity are discussed in Section~\ref{sec:threats-to-validity}, followed by conclusions and future directions in Section~\ref{sec:conclusions}.

    \section{Preliminaries}
    \label{sec:preliminaries}
    \subsection{Machine Learning}

    ML is a branch of artificial intelligence that focuses on creating systems that learn patterns from data and improve automatically with experience. In healthcare, ML supports diverse applications including disease identification, predictive diagnostics, image analysis, epidemic control, treatment optimization, genomic analysis, and surgical automation\cite{Qayyum2021}. These techniques enable researchers to analyse population-level data such as demographic patterns, symptom clusters, and disease correlations, classifying data based on features to derive insights for diagnosis, prediction, and analysis of diseases.

    ML approaches can be categorised into two main types: supervised and unsupervised learning. Supervised learning utilises labelled data, where input features are paired with corresponding outputs. The algorithm learns to predict outputs for new inputs by recognising patterns from training data. This approach is commonly employed for classification, regression, and object detection tasks. Key supervised algorithms include Logistic Regression~\cite{Hosmer2013}, Random Forests~\cite{Breiman2001}, and Naive Bayes~\cite{Hand2001}. In regression problems, the output variable represents a continuous value (e.g., 'pounds' or 'kilogrammes'), while classification problems involve categorical outputs (e.g., 'disease' or 'no disease'). Unsupervised learning, conversely, operates on unlabelled data without predefined outputs. These algorithms identify inherent patterns and relationships within datasets, making them suitable for clustering, dimensionality reduction, and anomaly detection. K-Means~\cite{JB1967} represents a prominent unsupervised learning technique used for grouping similar data points into clusters based on their characteristics.

    This work evaluates one unsupervised algorithm (K-Means) and three supervised methods (Random Forest, Logistic Regression, Naive Bayes). \\ [6pt]
    \emph{(i) K-means} is an unsupervised clustering algorithm that partitions data into $K$ distinct clusters by iteratively minimising the within-cluster sum-of-squares. The algorithm assigns data points to the nearest centroid and subsequently updates these centroids until convergence is achieved. In healthcare applications, K-means facilitates patient stratification by identifying cohorts with similar clinical characteristics, thereby enabling more targeted therapeutic interventions. This approach has demonstrated efficacy in medical image segmentation, particularly for tumour delineation, and in identifying latent patterns within complex clinical datasets that may elude conventional analytical methods.
    \\ [6pt]
    \emph{(ii) Logistic Regression} is a supervised learning algorithm that models the probability of a binary outcome through the logistic function, which maps the linear combination of predictors to a probability range $[0,1]$. This parametric approach is particularly valuable for medical diagnostics and prognostics, where quantifying the probability of disease occurrence is essential. The algorithm has been extensively applied in predicting the presence of chronic conditions such as diabetes, cardiovascular risk assessment, and clinical decision support. Its interpretability offers significant advantages in healthcare contexts where transparency in predictive modelling is paramount.
    \\ [6pt]
    \emph{(iii) Random Forest} represents an ensemble learning technique that constructs multiple decision trees during training and aggregates their predictions, typically through majority voting. This approach mitigates overfitting and enhances generalization, rendering it particularly suitable for both binary and multi-class classification challenges in medical contexts. Random Forest algorithms have demonstrated considerable utility in analysing complex medical datasets, including radiological images and electrocardiographic data, predicting disease progression, and identifying optimal therapeutic compounds based on molecular characteristics.
    \\ [6pt]
    \emph{(iv) Naive Bayes} classifiers apply Bayes' theorem with the naive assumption of conditional independence among features given the class label. This probabilistic approach is computationally efficient and performs remarkably well in various text classification tasks despite its simplifying assumptions. In medical contexts, Naive Bayes algorithms effectively process symptom-based diagnostic classification and medical document categorization. They have shown particular efficacy in diabetes detection when applied to clinical datasets, offering a balance between computational simplicity and predictive performance.


    \subsection{Differential Privacy}

    DP is a data sequestration technique that aims to safeguard individual privacy while enabling analysis of large datasets. This mathematical approach introduces controlled noise to data, making the outputs of queries differing by at most one record indistinguishable, thereby ensuring individual privacy \cite{Soria-Comas2017}. The DP guarantee for two datasets is given by:



    \begin{equation*}
        \mathrm{Pr}[\mathrm{M(D)}\in\mathrm{S}]\leq{e^\varepsilon}\mathrm{Pr}[\mathrm{M(D')}\in\mathrm{S}]
    \end{equation*}
    where:
    \begin{itemize}
        \item $M$: Randomised algorithm (i.e., $query(db) + noise$ or $query(db + noise)$)
        \item $S$: All potential outputs of M that could be predicted
        \item $D$: Entries in the database
        \item $D'$: Entries in adjacent database
        \item $\varepsilon$ : Privacy parameter that bounds the ratio of probabilities of obtaining the same output when the mechanism is run on adjacent databases
    \end{itemize}

    The databases $D$ and $D'$ differ by at most a single record, and the mechanism is differentially private if the results of $M(D)$ and $M(D')$ are almost indistinguishable for every choice of $D$ and $D'$. The $\varepsilon$ parameter quantifies the privacy guarantee. Rényi divergence~\cite{Asoodeh2021} is used in variants of DP to measure the difference between the probability distributions of the mechanism's outputs when applied to adjacent datasets.

    \subsubsection{Types of DP}
    \label{subsec:typesofdp}
    DP can be classified along two primary dimensions. First, based on implementation architecture, into Centralised DP and Local DP. Second, based on methodology, into Pure DP, Query-Level DP, and Moment DP~\cite{Abadi2016}.\\ [6pt]
    \emph{(i) Centralised DP} applies noise to a centralised database maintained by a trusted curator, protecting the privacy of the entire dataset while enabling useful analysis. The trusted curator manages both the raw data and the privacy-preserving mechanisms.\\  [6pt]
    \emph{(ii) Local DP} introduces noise directly to individual data points before their collection or aggregation, ensuring that sensitive information is protected at its source. This approach eliminates the need for a trusted curator as privacy guarantees are applied locally.\\ [6pt]
    \emph{(iii) Pure DP} provides mathematical guarantees that analysis results cannot reveal information about any individual record, regardless of the type of analysis performed or background knowledge available to potential adversaries.\\  [6pt]
    \emph{(iv) Query-level DP} applies privacy protections to specific queries rather than the entire dataset, allowing for fine-grained privacy budgeting based on query sensitivity and importance.\\ [6pt]
    \emph{(v) Moment DP} focuses on protecting statistical moments of the data distribution (e.g., mean, variance) by adding calibrated noise to these summary statistics rather than to raw data points, often resulting in improved utility for statistical analyses.

    \subsubsection{Noise Types in DP}
    DP achieves data protection by introducing calibrated noise to dataset elements or query results, preserving individual privacy while maintaining analytical utility. Various noise mechanisms can be employed, each with distinct statistical properties and application domains:\\ [6pt]
    \emph{(i) Laplace Noise Mechanism} applies noise drawn from the Laplace distribution, characterised by a symmetric probability density function centered at zero with exponentially decaying tails. This mechanism is widely implemented due to its mathematical tractability and strong theoretical guarantees for bounded-sensitivity functions, particularly in contexts requiring $\varepsilon$-DP with no additional parameters.\\ [6pt]
    \emph{(ii) Gaussian Noise Mechanism} introduces noise sampled from the Gaussian (normal) distribution, yielding the bell-shaped probability curve familiar in statistical applications. This approach provides $(\varepsilon, \delta)$-DP and typically demonstrates greater resilience to outliers than Laplace noise. The Gaussian mechanism is particularly effective for high-dimensional data and functions with $L2$-sensitivity constraints.\\ [6pt]
    \emph{(iii) Exponential Noise Mechanism} extends DP beyond numerical queries to selection problems by sampling outputs with probability exponentially proportional to their utility scores. This mechanism facilitates privacy-preserving selection from discrete sets and is widely employed for non-numerical applications such as identifying maxima or selecting optimal elements.\\ [6pt]
    \emph{(iv) Poisson Noise Mechanism} incorporates noise from the Poisson distribution, which models discrete count events. This approach is particularly well-suited for count queries, contingency tables, and histogram analyses where integral outputs are required. Its discrete nature makes it appropriate for applications involving event frequencies or integer-valued datasets.\\ [6pt]
    \emph{(v) Discrete Noise Mechanism} applies specially designed noise distributions to categorical data or discretised numerical values. This approach preserves the discrete structure of the underlying data while providing DP guarantees, making it particularly valuable for privacy-preserving analysis of categorical attributes and binned numerical data.

    \subsubsection{Underlying Concepts of DP}
    \noindent
    \emph{(i) Privacy Parameters}: DP establishes quantifiable boundaries regarding how much information about an individual's presence in a database may be disclosed to external parties. The parameters $\varepsilon$ and $\delta$ define these boundaries, characterising the privacy guarantees offered by a randomised privacy-preserving algorithm ($M$) applied to a specific database ($D$).
    \begin{itemize}
        \item \emph{Privacy Budget/Privacy Loss ($\varepsilon$)}: The $\varepsilon$ parameter quantifies the strength of privacy protection. A smaller $\varepsilon$ value corresponds to stronger privacy guarantees, whilst larger values indicate greater privacy loss and potentially more useful but less protected data.
        \item \emph{Probability to Fail/Probability of Error ($\delta$)}: The delta parameter accounts for the probability of privacy protection failure, specifically, the likelihood that a query might reveal an individual's presence in the dataset. Such failures may occur $\delta \times n$ times, where $n$ represents the number of records \cite{Abadi2016}.
    \end{itemize}
    \noindent
    \emph{(ii) Centralised Versus Local DP}:
    As discussed in Section~\ref{subsec:typesofdp}, two principal implementation strategies exist for DP. Centralized DP employs a trusted curator who applies precisely calibrated noise to query results. This approach typically utilises Laplace or Gaussian noise mechanisms and is commonly referred to as the trusted curator model. Conversely, local DP operates without requiring a trusted intermediary, hence its designation as the untrusted curator model. In local DP implementations, data undergoes randomization before curator access. A trusted entity may also employ local DP to simultaneously randomise all database records. Local DP algorithms frequently produce more heavily perturbed data, as noise is applied at the individual record level. This approach provides a particularly rigorous privacy standard with plausible deniability guarantees, establishing local DP as a leading methodology for privacy-preserving data collection and distribution, but at the cost of model utility.\\ [6pt]
    \noindent
    \emph{(iii) Correlated Sensitivity}:
    Whilst Global Sensitivity effectively measures the maximum number of correlated records, it fails to account for the degree of data correlation. The concept of Correlated Sensitivity addresses this limitation by quantifying the cascading impact on related records when a single record undergoes modification \cite{Zhang2020}. This refinement proves particularly valuable when analysing datasets containing interdependent entries, a common characteristic of complex relational databases.


    \subsubsection{Laplace Noise in Differential Privacy}



    The Laplace distribution's probability density function (PDF) is defined by the location parameter ($\mu$) and the scale parameter ($b$):

    \[
        f(x|\mu, b) = \frac{1}{2b} \exp\left(-\frac{|x - \mu|}{b}\right)
    \]

    where:
    \parskip 0pt
    \begin{itemize}
        \item $x$ is a random variable.
        \item $\mu$ represents the center of the distribution.
        \item $b$ controls the width of the distribution.
    \end{itemize}

    \begin{figure}[htpb]
        \centerline{\includegraphics[width=93mm,height=45mm,scale=1]{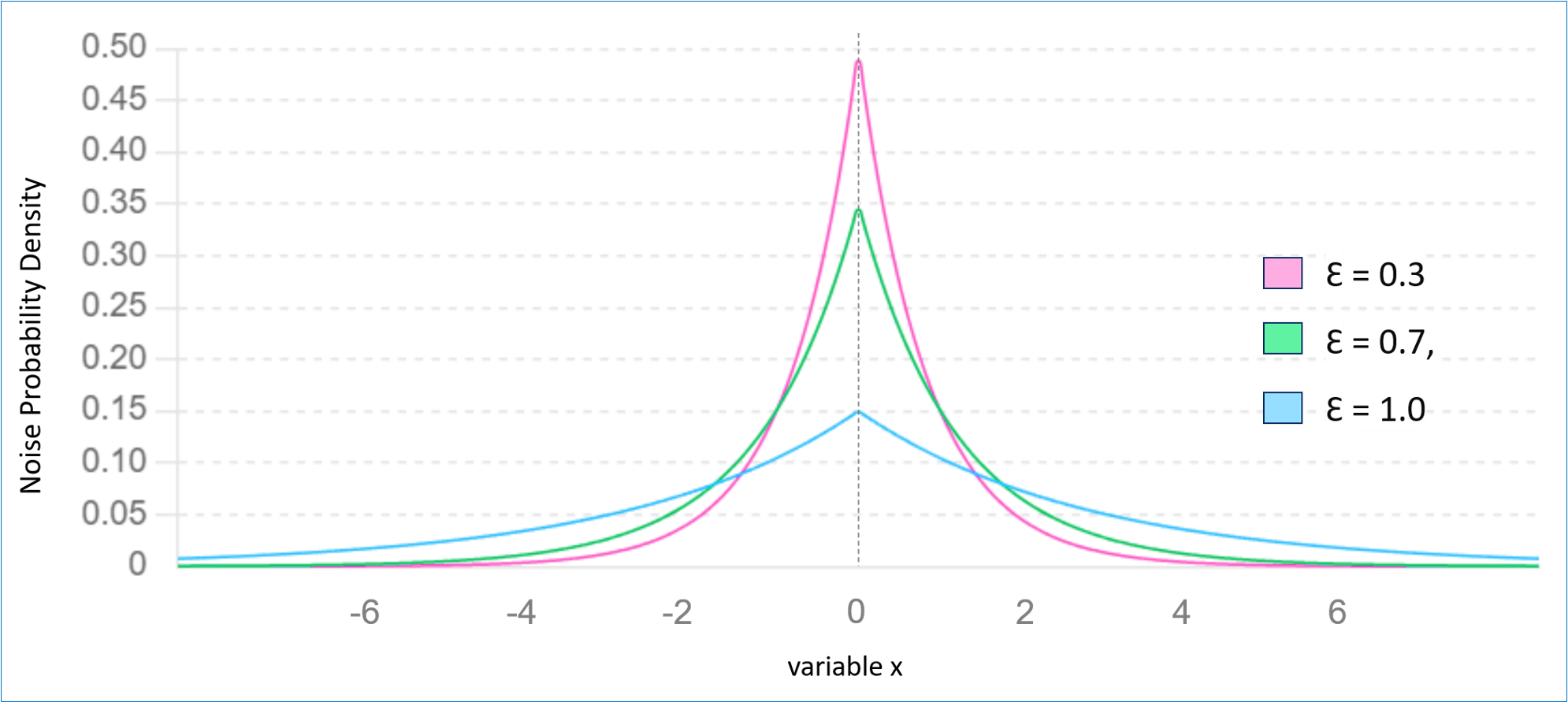}} 
        \caption{Laplace Noise Distribution for Different $\varepsilon$ Values} 
        \label{fig:laplace-noise-distribution} 
    \end{figure}

    In DP applications, this mechanism injects calibrated noise drawn from this distribution into query results. The noise calibration depends on the query sensitivity $\Delta f$ and the privacy budget parameter $\varepsilon$, as illustrated in Figure~\ref{fig:laplace-noise-distribution}.

    \begin{itemize}
        \item \emph{Sensitivity ($\Delta f$)}: The sensitivity parameter quantifies the maximum possible change in a query's output when a single record is either added to or removed from the dataset. For count queries with binary attributes, the sensitivity typically equals one, as an individual record can influence the count by at most one unit.

        \item \emph{Privacy Parameter ($\varepsilon$)}: The privacy budget parameter establishes the privacy-utility trade-off, with smaller values providing stronger privacy guarantees at the expense of analytical precision.
    \end{itemize}

    The Laplace mechanism determines the scale parameter ($b$) according to:

    \[
        b = \frac{\Delta f}{\varepsilon}
    \]

    This formulation ensures that the noise magnitude is proportional to the query sensitivity and inversely proportional to the privacy budget, thus maintaining mathematical guarantees of $\varepsilon$-DP.





    \subsubsection{Gaussian Noise in Differential Privacy}
    The Gaussian distribution's PDF is defined by the mean ($\mu$) and standard deviation ($\sigma$):

    \[
        f(x|\mu, \sigma) = \frac{1}{\sigma \sqrt2\pi} \exp\left(- \frac{|x - \mu|^2}{2\sigma^2}\right)
    \]

    where:
    \parskip 0pt
    \begin{itemize}
        \item $x$ is a random variable for which the probability density is calculated.
        \item $\mu$ represents the center of the distribution.
        \item \(\sigma\) is the standard deviation of the distribution, which controls the spread or width of the distribution.
    \end{itemize}

    In DP implementations, the Gaussian mechanism provides $(\varepsilon, \delta)$-DP, where $\delta$ represents a small probability of privacy failure. The mechanism is $(\varepsilon, \delta)$-differentially private when the standard deviation $\sigma$ of the Gaussian noise satisfies:

    \[
        \sigma = \sqrt{2log(1.25/\delta)}\frac{\Delta_2 f}{\varepsilon}
    \]

    where \(\varepsilon\) is the privacy parameter and \(\delta\) is the probability of failure.

    \begin{figure}[htpb]
        \centerline{\includegraphics[width=93mm,height=45mm,scale=1]{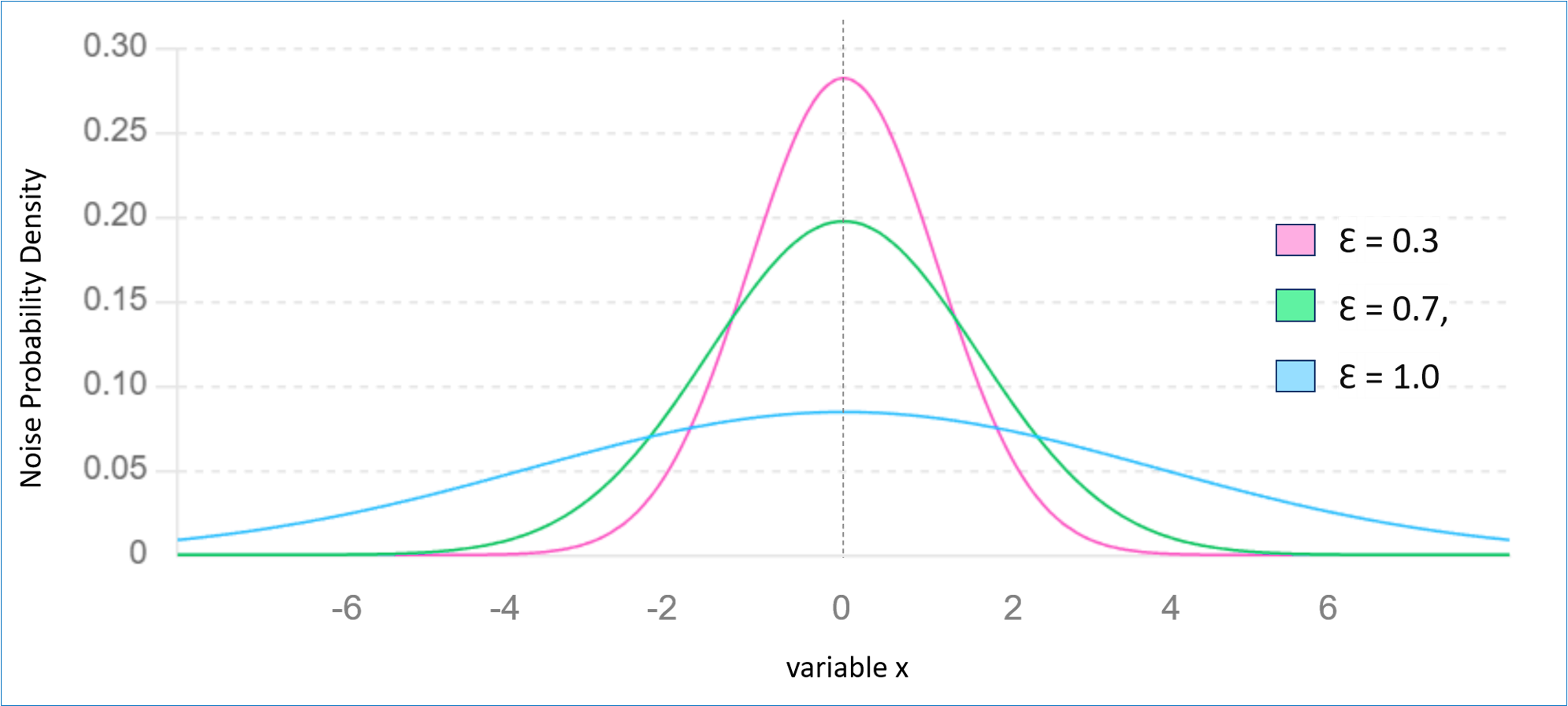}} 
        \caption{Gaussian Noise Distribution for Different $\varepsilon$ Values} 
        \label{fig:gaussian-noise-distribution} 
    \end{figure}

    As illustrated in Figure~\ref{fig:gaussian-noise-distribution}, the Gaussian distribution produces the characteristic bell-shaped curve, with its noise distribution varying based on the privacy parameter $\varepsilon$.

    \begin{itemize}
        \item \emph{Sensitivity ($\Delta f$)}: The Gaussian mechanism utilises $L2$ sensitivity (squared sensitivity), measuring the impact of changing a single data point through Euclidean distance. This approach is particularly effective for high-dimensional data analysis, mean calculations, and ML model training. For instance, in a dataset of $n$ entries ranging from $0$ to $1$, the $L2$ sensitivity equals $1/n$, as adding or removing one value alters the mean by at most $1/n$.

        \item \emph{Privacy Parameter ($\varepsilon$)}: As in the previous section, the privacy budget parameter establishes the trade-off between privacy protection and data utility.
    \end{itemize}

    \subsubsection{Applications of DP}
    Table~\ref{table:dp-applications} presents various domains that benefit from DP guarantees for statistical analysis and applications while preserving individual privacy. Key application areas include: \\ [6pt]
    \emph{(i) Healthcare}: DP enables secure sharing of electronic health records and clinical trial data, allowing researchers to extract insights whilst maintaining patient confidentiality \cite{Sun2022, Jiang2022}.\\ [6pt]
    \emph{(ii) Financial Services}: Financial institutions employ DP for fraud detection, risk assessment, and customer behaviour modelling. This protects individuals' financial information from exposure that could lead to identity theft or fraudulent activities \cite{Lin2021}.\\ [6pt]
    \emph{(iii) Social Media and Advertising}: Platforms implement DP to analyse user behaviour for personalised experiences. Google has applied this approach in its advertising platforms to create user profiles without exposing individual behaviours, addressing growing privacy concerns in digital advertising~\cite{Jiang2023, Lindell2011}.\\ [6pt]
    \emph{(iv) Transportation}: DP extracts insights from journey patterns and traffic flow data without revealing individual travel habits. This protection is essential as location data can reveal sensitive information about individuals' routines and behaviours \cite{Cai2019}.\\ [6pt]
    \emph{(v) Research and Academia}: DP supports research by enabling examination of sensitive datasets while preserving participant anonymity. This balance is crucial for maintaining ethical standards in research whilst facilitating valuable data-driven discoveries \cite{Zhan2024}.\\[6pt]
    \emph{(vi) Government and Policy-making}: DP facilitates evidence-based policy development using large-scale datasets. This protection is necessary as government data often contains comprehensive citizen information that, if compromised, could lead to significant privacy violations \cite{Saeidian2021}.\\ [6pt]
    \emph{(vii) Web Browsing Analysis:} Google implemented DP techniques through their RAPPOR framework for Chrome browsing data to identify performance issues such as slow-loading web pages. This approach enables system improvements whilst protecting individual browsing patterns from surveillance or profiling~\cite{Erlingsson2014}.\\[6pt]
    \emph{(viii) Location-Based Services:} Google Maps utilises DP to provide real-time traffic updates and location-based services from millions of users' data. This implementation preserves user anonymity whilst delivering accurate collective insights about traffic conditions and business popularity~\cite{google2019}.\\[6pt]
    \emph{(ix) Search Query Analysis:} Google's search engine employs DP to enhance recommendations by analysing queries and user interactions. This enables improved search results without compromising individual search histories, which may contain sensitive personal information~\cite{Lindell2011}.\\[6pt]
    \emph{(x) Public Health Monitoring:} During the COVID-19 pandemic, Google released aggregated mobility reports using DP to protect individual identities. These reports assisted health authorities in understanding social distancing patterns whilst maintaining location privacy of individual users~\cite{Aktay2020}.

    \subsection{Blockchain}
    Blockchain technology represents a distributed, decentralised ledger architecture designed to record transactions across multiple computing nodes. The architecture comprises multiple interconnected blocks, with each block incorporating cryptographic references to its predecessor. This structure ensures that records cannot be altered or manipulated without modifying all subsequent blocks in the chain. Once data is entered onto the blockchain, it becomes immutable. This immutability confers strong resistance to tampering attempts by malicious actors~\cite{Jiang2022}.

    Several key features make blockchain particularly valuable for healthcare applications:
    \begin{itemize}
        \item \emph{Immutability:} Patient data, once stored on the blockchain, cannot be altered or tampered with, ensuring the integrity of medical records.
        \item \emph{Time-stamping:} All transactions are chronologically recorded with precise timestamps, providing an auditable history of healthcare events.
        \item \emph{Traceability:} Patient-related data, including hospitalizations and treatments, can be traced and categorized by geographical location and temporal parameters.
        \item \emph{Transparency:} All transactions are transparently viewable by authorised stakeholders within the healthcare ecosystem, facilitating improved coordination of care.
        \item \emph{Automation:} Smart contracts, which are self-executing protocols stored on the blockchain, automatically implement predefined actions when specific conditions are met without requiring intermediary involvement. These programmable agreements enable automated healthcare workflows such as insurance claim processing, clinical trial management, and medication supply chain verification.
    \end{itemize}
    These features collectively enhance the reliability and security of medical data systems~\cite{Rachakonda2021}. Blockchain's immutable architecture provides an effective framework for regulatory compliance with health authority mandates while enabling secure information exchange among healthcare professionals. This technology is particularly valuable in healthcare contexts characterised by distributed trust requirements, where multiple stakeholders must verify data integrity and provenance without centralised authority. The resultant interoperability across disparate systems addresses a significant challenge in contemporary healthcare information management.

    \begin{table*}[htbp]
        \centering
        \caption{Applications of Differential Privacy}
        \label{table:dp-applications}
        \scriptsize
        \begin{tabular}{|p{3cm}| p{6cm}| p{8cm}|}
            \hline

            \textbf{Application} & \textbf{Description} & \textbf{Factors to Consider} \\
            \hline
            Data Analysis
            & Aggregating and analyzing sensitive data & - Privacy budget (\(\varepsilon\)) for the desired level of privacy\\
            & & - Noise mechanism (e.g., Laplace, Gaussian) based on data distribution\\
            & & - Data sensitivity and scale\\
            & & - Query types and frequency\\
            \hline
            Recommendation Systems & Personalized recommendations & - Protecting user preferences while providing personalized suggestions\\
            & & - Trade-off between utility and privacy\\
            & & - Ensuring diversity in recommendations\\
            \hline
            Healthcare Research & Medical data analysis and research & - Compliance with healthcare regulations (e.g., HIPAA)\\
            & & - Preserving patient privacy while conducting studies\\
            & & - Anonymization techniques for sharing medical records\\
            \hline
            Location Privacy & Protecting user location data & - Adding noise to location data for anonymity\\
            & & - Balancing location accuracy with privacy guarantees\\
            & & - Considering adversarial attacks on location data\\
            \hline
            Social Media Analysis & Analyzing social media posts and trends & - Anonymizing user data to protect identity\\
            & & - Maintaining the ability to detect trends and sentiment\\
            & & - Privacy implications of social network analysis\\
            \hline
            Census Data & Collecting and sharing demographic information & - Protecting individuals' privacy in census data\\
            & & - Balancing accuracy of demographic data with privacy guarantees\\
            & & - Adherence to legal and ethical guidelines\\
            \hline
            Smart Cities & Analyzing data from IoT sensors for urban planning & - Anonymizing sensor data to protect privacy\\
            & & - Ensuring data integrity and security\\
            & & - Handling diverse data sources and formats\\
            & & - Public awareness and consent\\
            \hline
            Finance and Banking & Protecting financial transaction data & - Compliance with financial regulations (e.g., GDPR)\\
            & & - Ensuring transaction privacy\\
            & & - Detecting fraud while preserving customer privacy\\
            & & - Secure data sharing mechanisms\\
            \hline
            Educational Research  & Analyzing student performance and learning outcomes & - Balancing the need for research with student privacy\\
            & & - Consent and data sharing agreements\\
            & & - Anonymizing student data to prevent identification\\
            & & - Ethical considerations\\
            \hline
            Transportation & Analyzing traffic patterns and congestion & - Protecting user privacy while analyzing traffic data\\
            & & - Noise addition to GPS data for anonymity\\
            & & - Data aggregation techniques for traffic analysis\\
            & & - Privacy-aware routing algorithms\\
            \hline

        \end{tabular}
    \end{table*}


    \subsection{Different layers of Computing Architecture}

    The evolution of computing paradigms has facilitated a hierarchical approach to processing healthcare data, addressing various requirements including latency, resource constraints and data volume. This section examines the distinctive characteristics of each architectural layer and their specific healthcare applications.
    \\ [6pt]
    \begin{enumerate} [(i) ]
    \item \emph{Edge Computing}:
    Edge nodes provide computational offloading, storage and caching capabilities for IoT management, facilitating reduced computation requirements and rapid response times for time-sensitive applications \cite{Kumar2020}.
    In healthcare contexts, edge computing serves as an intermediary processing layer, managing data that requires swift analysis but can tolerate minimal processing delays. A remote healthcare facility might deploy edge servers to process and store patient vital signs such as blood pressure, oxygen saturation and cardiac rhythms. Medical practitioners can examine this locally processed data to make timely clinical decisions, while selectively determining which information warrants transmission to cloud infrastructure for comprehensive analysis and long-term storage.\\
    \item \emph{Cloud Computing}: Cloud computing provides on-demand network access to a shared pool of configurable computing resources, including servers, storage, applications and databases \cite{Yan2021}. This model centralises data processing and storage on remote server infrastructure, offering significant computational capacity and storage volumes at the expense of increased latency compared to edge-oriented paradigms.
    While not optimised for time-critical healthcare functions, cloud computing excels in scenarios requiring substantial resources for complex analysis or extensive data retention. Healthcare organizations leverage cloud platforms to maintain comprehensive historical patient records, conduct longitudinal research and perform sophisticated analytical operations. Cloud-based solutions securely store EHRs from multiple healthcare institutions, enabling researchers and clinicians to identify long-term health trends, population-level insights and evidence-based treatment protocols through analysis of aggregated datasets.
    \end{enumerate}

    \section{Literature Review}
    \label{sec:literature-review}

    Recent research has explored various approaches to implementing DP in IoT-Cloud systems. Sun et al. \cite{Sun2020} examined the relationship between inference and data privacy, comparing DP implementations using Bayes probability and Gauss-Seidel methods for enhanced data privacy, scalability and reduced network latency. Their work highlighted the need for further investigation regarding the impact of communication delays and packet losses on accuracy and privacy in IoT environments.

    While addressing communication challenges in IoT networks, researchers have explored blockchain integration with DP mechanisms, despite blockchain's own consensus overhead contributing to communication delays. Jia et al. \cite{Jia2022} implemented a blockchain-enabled federated data protection aggregation scheme with DP, utilising K-means clustering with Adaptive Boosting and Laplacian noise to improve security across multiple datasets. Their research indicated that homomorphic encryption techniques could further enhance secure data exchange in Industrial IoT (IIoT) contexts. Similarly, Gai et al. \cite{Gai2020} proposed a blockchain-based Internet of Edge model incorporating Q-Learning algorithms with Laplace noise distribution to implement tamper-resistant DP, though they noted high energy consumption as a limitation requiring future work.

    Several studies have investigated noise distribution mechanisms for DP. Hu et al. \cite{Hu2020} implemented Federated Learning DP using Gaussian methods, demonstrating that heterogeneous noise perturbation achieves robust accuracy suitable for real-world scenarios. Complementing this work, Cai et al. \cite{Cai2022} developed a multimodal DP framework for local DP that demonstrated the flexibility of Gaussian noise in balancing privacy protection with high data utility. Chowdhury et al. \cite{RoyChowdhury2020} described a DP framework utilising a noisy max algorithm with Laplace mechanism that provided accuracy guarantees, suggesting potential for practical application with extensive empirical evaluation on real-world datasets.

    Foundational work by Dwork \cite{Dwork2008} presented various privacy-preserving data analysis techniques established since the early 2000s, providing systematic approaches to achieving privacy protection. Taking a different architectural approach, Bi et al. \cite{Bi2022} designed a privacy-isolation zone where sensitive user information is removed from data collected at the IoT endpoints before transmission to cloud systems for analytics.

    Table~\ref{table:lit-review} summarises notable recent efforts in guaranteeing privacy protection for sensitive datasets, highlighting the diversity of approaches and their respective strengths in addressing the privacy-utility trade-off in IoT-Cloud systems.

    \begin{table*}[htbp]
        \centering
        \scriptsize
        \caption{DP Literature Review}
        \label{table:lit-review}
        \begin{tabular}{|p{3.5cm}| p{4.5cm}| p{5cm}| p{4cm}|}
            \hline

            \textbf{Author} & \textbf{Algorithm} & \textbf{Performance} & \textbf{Research Gaps}\\
            \textbf{Concept/Model} & \textbf{Implementation} & \textbf{Advantages} &  \textbf{Future Work}\\
            \hline

            Huang et al.,\cite{Huang2020} 2020, DP-ADMM:ADMM-Based Distributed Learning with DP &
            - DP-Alternating direction method of multipliers (DP-ADMM) \newline
            - Real-world dataset: Adult data set from UCI ML Repository \newline
            - Gaussian noise &
            - Distributed Learning \newline
            - Noise-resilient, convergent, High privacy guarantee  &
            Increased computation time and sensitive to hyper parameters\\
            \hline

            Saeidian et al.,\cite{Saeidian2021} 2021, Quantifying Membership Privacy via Information Leakage &
            - Private Aggregation of Teacher Ensembles (PATE) framework \newline
            - Public data set \newline
            - Laplace distributed noise  &
            - Novel framework for measuring membership privacy \newline
            - Accurate measure of membership privacy &
            Difficult to implement in practice it make challenges for organization to apply framework in real world situation \\
            \hline

            Gai et al.,\cite{Gai2020} 2020, DP-Based Blockchain for Industrial Internet-of-Things &
            - Blockchain Internet of Edge model \newline
            - Q-learning algorithm \newline
            - Laplace's distribution noise &

            - BIoE model enhance the Privacy-Preserving capacity \newline
            - Tamper resistant &
            Study impact of data and noise on energy cost in future works \\
            \hline

            Li et al.,\cite{Li2023} 2023, Optimal Trading Mechanism Based on DP and Stackelberg Game in Big Data  &
            - Optimized Unary Encoding (OUE) \newline
            - LDP based gradient iteration (LGI) algorithm &
            Guarantee both Privacy and utility for data provider and the users &
            Optimizing our model with more measured data from data platform and optimizing game theory using Reinforcement learning theory \\
            \hline

            Zhang et al.,\cite{Zhang2020} 2019, Correlated DP: Feature Selection in ML &
            - CR-FS Scheme \newline
            - Mean absolute error (MAE) \newline
            - ML algorithm  SVM and LR \newline
            - Data sets: Adult, Breast Cancer, Titanic, Porto Seguro
            &
            - Better prediction results with ML tasks \newline
            - Better trade-off between data utility and privacy leaks \newline
            - Reduce Data correlation
            &
            Data correlation may bring new errors with different queries \\
            \hline

            Bi Jia et al., \cite{Jia2022}, 2022, Federated Learning Data
            Aggregation with DP, Homomorphic Encryption and  Blockchain in IIoT &
            - K-means clustering with DP, Homomorphic encryption
            - Random Forest, AdaBoost &
            Improved F1-score and accuracy compared normal K-Means &
            Test for secure exchange and sharing for Enterprise IIoT\\
            \hline

            Zhang et al., \cite{Zhang2023APDP} 2023, APDP: Attack-Proof Personalized DP Model for  Smart Homes &
            - APDP model \newline
            - Fog computing \newline
            - Real-world Smart home data set
            &
            - Improved performance with enhanced bandwidth and reduced service latency \newline
            - Defeat collusion attack under multiple circumstances \newline
            - Achieved optimize trade-off between Privacy protection and data utility
            &
            Explore cross discipline techniques to further optimize the trade-offs \\
            \hline

            Zhang et al., \cite{Zhang2022DPAuction} 2022, DP-Based double Auction for data market in Blockchain IoT  &
            - Double-Auction Normal Transaction Method (DANTM) \newline
            - Double-Auction Transaction Method Based on DP (DADPM) \newline
            - Gaussian mechanism &
            - Protects participants bid information \newline
            - Good truthfulness, Privacy, performance
            &
            Difficult to determine the size of the Noise \\
            \hline

            Ali et al., \cite{Ali2022} 2022, A privacy enhancing model
            for IoT using three-way decisions and DP &
            - Attribute Division Algorithm for DP (3WADD) \newline
            - Laplace noise \newline
            - Data sets: Titanic, Adult, Bank, Marketing, Heart Disease, Student Performance
            &
            - Automatic division of attributes for DP \newline
            - Information content and stability of data set
            &
            More sophisticated method three way decision attributes of IoT to motivate more organizations to use IoT \\
            \hline

            Miao Du et al., \cite{Du2018}, 2018, DP of Training Model
            in Wireless Big Data with Edge Computing &
            Laplace mechanisms with Output Perturbation (OPP) and Objective Perturbation (OJP) &
            Effectively privacy protection of  training data ensuring \(>95\%\) accuracy and data utility &
            Apply for practical usecases \\
            \hline

        \end{tabular}
    \end{table*}

    \parskip 0pt
    \section{Problem Statement}
    \label{sec:problem-statement}
    Medical records containing patient histories, diagnoses, treatments, and health-related information represent sensitive data requiring careful management. The central challenge in healthcare IoT-Cloud systems is balancing immediate emergency response capabilities with secure data handling that enables valuable research without compromising patient privacy. This research addresses the tension between processing IoT data with minimal latency for emergency response while simultaneously ensuring secure transmission of this data to cloud storage for long-term preservation, diagnosis, and analytics.\\ [6pt]
    \emph{Objectives:}\\
    (i) Enhance the response speed of IoT-Cloud healthcare systems for emergency scenarios \\
    (ii) Provide robust privacy guarantees for patient data while maintaining research utility \\
    (iii) Ensure reliable and secure storage of healthcare transaction data in cloud environments \\

    \noindent
    \emph{Constraints:}\\
    The fundamental constraint governing this research is the necessity to maintain a balance between robust patient privacy protection and preserving sufficient data utility for scientific inquiry. This represents the classical privacy-utility trade-off that requires careful optimization within the proposed framework to ensure both ethical data handling and meaningful analytical capabilities.

    \section{Threat Model and Security Analysis}
    \label{sec:threat-model}

    This section establishes a comprehensive threat model for healthcare IoT-Cloud systems, defining adversary capabilities, attack surfaces, and security objectives that our proposed solution must address.

    \subsection{System Model and Trust Assumptions}

    Healthcare IoT-Cloud systems comprise three hierarchical computational tiers: IoT devices, Edge computing infrastructure and Cloud data centers as shown in Figure~\ref{fig:multi-layered-architecture}. Each tier presents distinct trust characteristics and security vulnerabilities. IoT devices including patient wearables and medical sensors operate under direct patient or healthcare provider control, typically with physical security measures limiting unauthorized access. However, these devices often possess limited computational resources for implementing sophisticated security protocols.

    Edge computing nodes occupy intermediate positions in the architectural hierarchy, processing data from multiple IoT sources while forwarding aggregated information to cloud infrastructure. These nodes may be operated by third-party service providers or healthcare organizations, introducing varied trust assumptions. The computational resources available at these layers enable more sophisticated processing but also present larger attack surfaces for potential adversaries. Cloud infrastructure, while offering substantial computational and storage capabilities, operates under the control of external service providers and may be accessible to multiple stakeholders including researchers, pharmaceutical companies, and insurance providers, each with potentially conflicting interests regarding data access and privacy.

    In distributed ledger implementations for healthcare data integrity, we must consider the trust model of consensus participants. Permissioned blockchain networks restrict participation to authorized entities, whereas public blockchains allow arbitrary participation. The security guarantees of these systems depend fundamentally on assumptions about the proportion of honest versus malicious consensus participants. Byzantine fault tolerance models typically assume that fewer than one-third of participants exhibit arbitrary malicious behavior, while other consensus mechanisms may require different trust assumptions.

    \subsection{Adversary Model}

    We classify potential adversaries into three categories with progressively increasing capabilities, reflecting the diverse threat landscape confronting healthcare information systems.

    \subsubsection{Type I: Passive Adversary (Honest-but-Curious)}

    The Type I adversary represents entities that faithfully execute system protocols but attempt to extract sensitive information through observation and analysis. Such adversaries possess the capability to observe all data passing through compromised nodes within their control, including encrypted traffic patterns, query sequences, and aggregate statistical outputs. They maintain access to auxiliary information sources such as public demographic databases, voter registration records, medical literature, and previously published health statistics. These adversaries possess unlimited computational resources for offline cryptanalysis and statistical inference attacks, enabling them to perform sophisticated correlation analyses and apply advanced ML techniques to infer sensitive attributes.

    The Type I adversary can execute multiple queries on databases or trained ML models, potentially crafting query sequences designed to maximize information extraction while remaining within nominal system usage patterns. However, this adversary class is constrained by its adherence to system protocols: it cannot modify data in transit or at rest, cannot corrupt other system components beyond those under its direct control, and cannot forge cryptographic signatures or break established cryptographic primitives.

    The primary objectives of Type I adversaries include re-identification attacks, where anonymized medical records are linked to specific individuals through correlation with auxiliary information sources and exploitation of quasi-identifiers such as birthdate, postal code, and demographic attributes. Attribute inference represents another critical threat, wherein adversaries deduce sensitive health attributes including disease status, genetic predispositions, medication regimens, or behavioral risk factors for individuals of interest. Membership inference attacks attempt to determine whether a specific individual's data was included in a training dataset or statistical database, potentially revealing participation in sensitive medical studies or presence of stigmatized conditions.

    \subsubsection{Type II: Active Adversary (Malicious)}

    The Type II adversary extends beyond passive observation to actively manipulating system components and data flows. This adversary class possesses all capabilities of Type I adversaries while additionally being able to modify, inject, or delete data packets during network transmission. Active adversaries can compromise and assume control of Edge computing nodes, potentially affecting data processing for multiple IoT sources. They can submit carefully crafted queries specifically designed to amplify privacy leakage beyond what would be revealed through legitimate usage patterns, exploiting potential vulnerabilities in privacy protection mechanisms.

    Collusion represents a significant threat multiplier for Type II adversaries, as multiple compromised entities may pool their observations and capabilities to breach privacy or integrity protections that would withstand individual attackers. However, several constraints limit Type II adversary capabilities. Physical security measures protect IoT devices and Edge nodes from direct compromise in most deployment scenarios. Established cryptographic primitives including digital signature schemes, hash functions, and encryption algorithms are assumed to remain computationally infeasible to break. Distributed consensus mechanisms in blockchain implementations impose constraints on the ability of adversaries to unilaterally modify ledger contents.

    The attack objectives for Type II adversaries encompass data tampering, wherein patient medical records are modified to influence clinical decisions, insurance claim determinations, or legal proceedings. Privacy budget exhaustion attacks involve carefully sequenced queries designed to deplete privacy protection mechanisms, enabling subsequent queries to extract sensitive information with reduced protection. In distributed ML scenarios, gradient leakage attacks attempt to reconstruct training data from model parameter updates or gradient information exchanged during collaborative learning. Model poisoning attacks inject malicious training examples designed to corrupt the behavior of ML models, potentially causing misdiagnosis or inappropriate treatment recommendations.

    \subsubsection{Type III: Insider Adversary}

    The Type III adversary represents perhaps the most challenging threat class: authorized system users who abuse their legitimate access privileges for malicious purposes. Insider adversaries possess all capabilities of Type II adversaries while additionally holding valid system credentials and authorized access to various system components. These adversaries can access raw, unprotected data before privacy-preserving transformations are applied, possess detailed knowledge of system architecture, implementation details, and security mechanisms, and may abuse privileged access to bypass certain security controls intended to constrain external adversaries.

    Despite these extensive capabilities, insider adversaries face several constraints. System activity logging mechanisms record access patterns and operations performed by authenticated users. Role-based access control systems restrict the scope of data and operations accessible even to privileged users, implementing principle of least privilege and separation of duties. In systems employing cryptographic audit trails such as blockchain-based logging, insiders cannot forge or repudiate transactions without detection, as each operation is cryptographically signed and timestamped.

    Insider adversaries pursue several attack objectives. Privilege escalation attacks attempt to access data or perform operations beyond the adversary's authorized scope, potentially by exploiting software vulnerabilities or social engineering. Unauthorized disclosure involves exfiltration of protected health information for purposes including financial gain through sale to data brokers, competitive intelligence, blackmail, or personal curiosity. Cover-up attacks attempt to delete or modify audit logs to conceal prior malicious activities, though cryptographic logging mechanisms may render such attempts detectable or impossible.

    \subsection{Attack Surfaces and Threat Vectors}

    The multi-layer architecture of healthcare IoT-Cloud systems presents distinct attack surfaces at each computational tier, with threat vectors exploiting various system vulnerabilities.

    At the IoT and Edge computing layers, physical device compromise represents a primary threat vector. Adversaries may gain unauthorized physical access to wearable medical devices or home healthcare equipment, potentially extracting cryptographic keys, implanting malware, or tampering with sensor readings. Devices with limited computational resources may lack sophisticated security features such as secure enclaves or hardware-based attestation, making them vulnerable to firmware modification. The resource constraints of IoT devices also render them susceptible to denial-of-service attacks that exhaust battery power or overwhelm processing capacity.

    The Edge computing layer faces threats from network traffic analysis. Adversaries monitoring network communications may perform traffic analysis attacks, correlating encrypted packet sizes, timing patterns, and communication frequencies to infer sensitive information about patient conditions or healthcare activities even without decrypting packet contents. These intermediate layers process data from multiple IoT sources, presenting opportunities for cross-patient correlation attacks if data from different individuals is not properly isolated.

    A particularly significant threat across both Edge and Cloud layers involves inference attacks on ML models and aggregate statistics. Membership inference attacks analyse the behavior of trained models to determine whether specific individuals' data was included in training datasets, potentially revealing participation in sensitive medical studies. Attribute inference exploits correlations in released statistics or model predictions to deduce sensitive attributes that were not directly disclosed. Model inversion attacks attempt to reconstruct training data from model parameters or predictions, potentially recovering sensitive medical records. Property inference attacks deduce aggregate properties of training data that were not intended for release, such as prevalence of specific conditions or demographic correlations.

    Data integrity threats emerge at the Cloud storage layer. Adversaries with write access to storage systems may modify patient records to corrupt medical decision-making, fraudulently alter insurance claims, or manipulate research datasets. More subtle integrity violations involve selective deletion of records to bias statistical analyses or hide evidence of medical errors. The temporal dimension presents additional vulnerabilities, as adversaries may attempt to manipulate timestamps to obscure the chronology of medical events or treatment decisions.

    For systems employing distributed ledger technology to ensure data integrity, consensus mechanisms themselves present attack surfaces. Majority attacks in proof-of-work or proof-of-stake systems involve adversaries controlling sufficient computational power or stake to override consensus and modify ledger contents. Selfish mining strategies allow adversaries to gain disproportionate influence over consensus by strategically withholding and releasing blocks. Eclipse attacks isolate specific nodes from the honest network, feeding them false information about the state of the distributed ledger.

    Cross-layer threats span multiple architectural tiers. Linkage attacks combine information from multiple queries, potentially issued to different system components or at different times, to circumvent privacy protections that would be effective against individual queries. Collusion attacks involve multiple adversaries, potentially with compromised nodes at different architectural layers, pooling their observations to amplify information extraction. Composition attacks exploit the accumulation of privacy loss across multiple privacy-preserving mechanisms or repeated queries to the same underlying dataset.

    \subsection{Security Objectives}

    The threat landscape outlined above necessitates five fundamental security objectives that healthcare IoT-Cloud systems must satisfy.

    \textbf{Confidentiality} requires that individual patient records remain confidential even when aggregate statistics, trained ML models, or research findings derived from the data are publicly released. This objective extends beyond simple access control to require that released information provably limits what can be inferred about individuals. Quantifiable privacy guarantees must bound the maximum information leakage that can occur through any sequence of queries or analyses, even when adversaries possess arbitrary auxiliary information and unlimited computational resources for inference attacks.

    \textbf{Integrity} demands protection of patient data, ML models, audit logs, and system configurations against unauthorized modification. Any tampering attempts must be detectable through cryptographic verification mechanisms, and the system must maintain evidence of data provenance enabling verification of authenticity. Integrity protections must be tamper-evident, meaning that modifications leave detectable traces even if the adversary controls storage infrastructure. For critical healthcare data, integrity requirements may extend to non-repudiation, ensuring that entities cannot deny having performed specific operations.

    \textbf{Availability} ensures that healthcare services remain operational despite adversarial disruption attempts. This encompasses resilience against denial-of-service attacks, infrastructure failures, and resource exhaustion. For emergency healthcare scenarios, availability requirements include strict latency bounds on critical operations, as delays in responding to physiological emergencies can result in patient harm or mortality. Availability must be maintained even under partial system compromise, requiring redundancy and graceful degradation of service quality rather than complete failure.

    \textbf{Accountability} requires that all data access and modification operations be attributable to specific entities through immutable audit trails. Authenticated users must not be able to perform operations anonymously or repudiate actions they have taken. Audit mechanisms must themselves resist tampering, as adversaries may attempt to cover their tracks by modifying logs. Accountability enables forensic investigation of security incidents, deters insider attacks through the threat of detection, and supports regulatory compliance with healthcare privacy regulations requiring access logging.

    \textbf{Privacy Preservation} mandates formal, quantifiable guarantees limiting information leakage across multiple queries and analyses. Unlike confidentiality, which focuses on access control, privacy preservation addresses the fundamental tension between data utility and individual privacy in statistical databases and ML systems. Privacy guarantees must hold under worst-case adversarial behavior, accounting for arbitrary auxiliary information and composition of multiple analyses. The quantification of privacy loss must enable healthcare organizations to make informed decisions about acceptable risk levels while complying with regulatory requirements such as HIPAA in the United States or GDPR in the European Union.

    \section{Proposed Solution}
    \label{sec:proposed-solution}
    The threat model established in Section~\ref{sec:threat-model} identifies three adversary classes (honest-but-curious, malicious, and insider threats) along with attack surfaces spanning inference attacks, data tampering, and consensus attacks. To address this comprehensive threat landscape, our research proposes an integrated framework comprising three complementary components that provide defense-in-depth security:\\
    (i) A multi-layered IoT-Edge-Cloud computing architecture to improve response time for healthcare applications through strategic workload distribution \\
    (ii) Privacy-preserving DP techniques to maintain data utility for medical research whilst protecting patient identity, specifically addressing the inference attacks (re-identification, membership inference, attribute inference) posed by Type I and Type II adversaries \\
    (iii) Secure cloud storage utilising blockchain technology for healthcare transaction data, providing tamper-evident audit trails and integrity verification to counter data tampering threats from Type II and Type III adversaries

    To facilitate real-time processing for delay-intolerant emergency healthcare responses, the proposed architecture incorporates multiple computational layers as depicted in Fig.~\ref{fig:multi-layered-architecture}. This IoT, Edge and Cloud (IEC) framework is characterized by increasing compute capacity and storage permanence in the upper layers, counterbalanced by corresponding increases in communication latency. This design enables task distribution based on response urgency, with critical operations deployed to lower layers and persistent data directed to cloud infrastructure. The hierarchical nature of this architecture also implements defense-in-depth against the cross-layer threats identified in Section~\ref{sec:threat-model}, limiting the scope of compromise at each tier through role-based access control and computational isolation.

    We achieve privacy protection by applying DP techniques during the analysis phase, where systems are most vulnerable to compromise. As established in our threat model, Type I adversaries with unlimited computational resources and access to auxiliary information can perform sophisticated inference attacks through multiple queries on aggregated datasets to extract personal patient information. By applying calibrated noise to ML training datasets, we provide mathematical guarantees bounding information leakage even under worst-case adversarial conditions. For any two datasets $D$ and $D'$ differing by a single record, the probability ratio of observing any output is bounded by $\varepsilon$, ensuring that healthcare institutions can share data with researchers whilst preventing individual patient re-identification, thus preserving population-level insights for medical advancement.

    \subsection{Multi-Layered Compute-Storage Architecture}

\begin{figure}[htpb]
        \centerline{\includegraphics[width=90mm,scale=0.5]{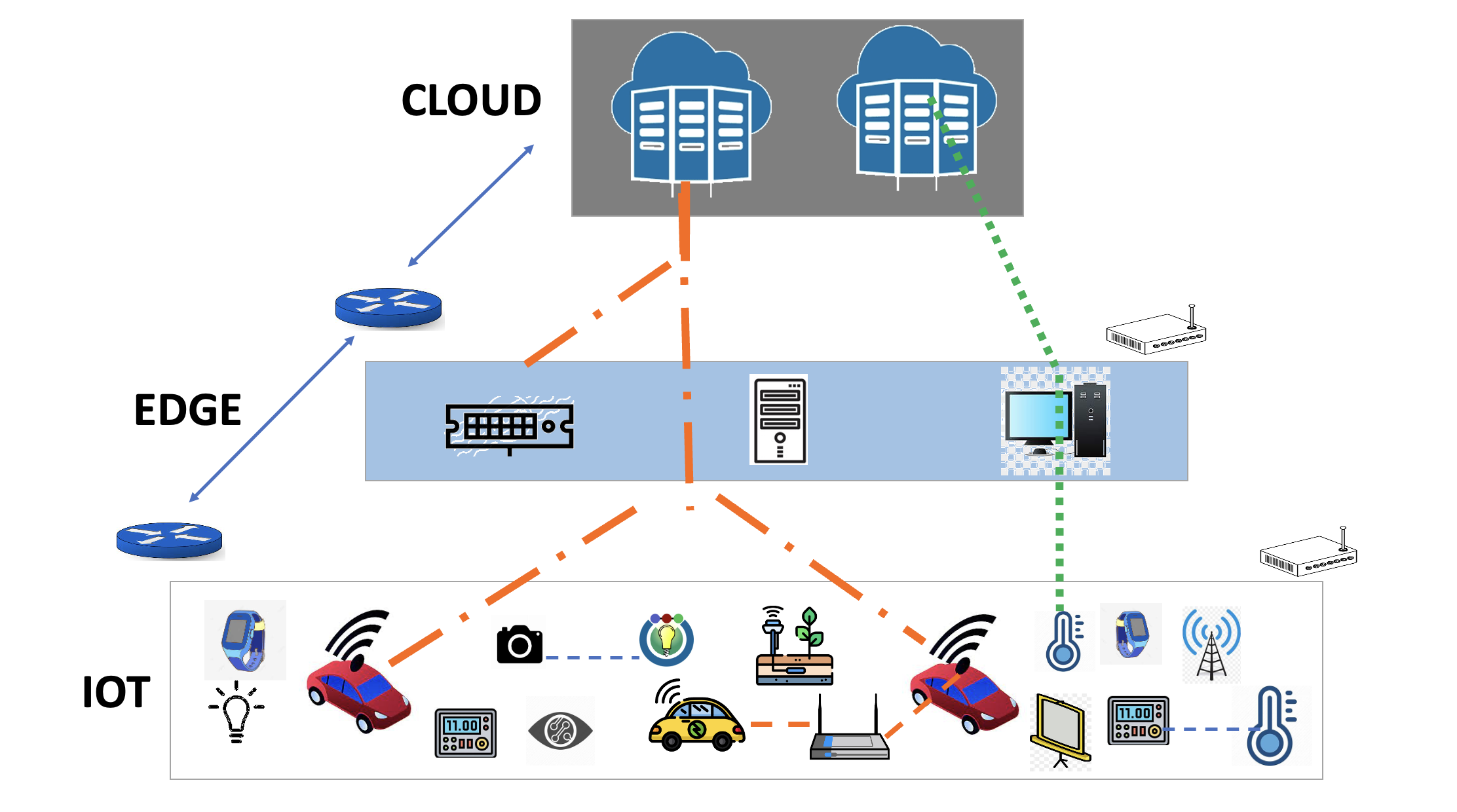}} 
        \caption{Multi-Layered Compute-Storage Architecture } 
        \label{fig:multi-layered-architecture} 
    \end{figure}

    The proposed solution incorporates a hierarchical edge and cloud architecture to ensure rapid response times and enhanced security for healthcare IoT applications, as illustrated in Fig.~\ref{fig:multi-layered-architecture}.

    The IEC architecture strategically distributes computational workloads and data based on proximity to data sources, processing speed requirements, and storage permanence needs. This distribution optimizes system performance for various healthcare scenarios with different response criticality profiles.

    The effectiveness of this approach is demonstrated through several healthcare applications. In emergency response scenarios, patient wearables detecting physiological abnormalities trigger immediate data processing at proximal Edge Computing nodes, while simultaneously transmitting data to Cloud storage via other Edge nodes for comprehensive analytics. This dual-path approach ensures both immediate intervention and long-term data utilization for population health insights by public health authorities. Child Health Information systems represent intermediate computational requirements, utilizing Edge nodes to compare current sensor readings against historical baselines for generating caregiver recommendations. Telemedicine applications employ Edge Computing resources for processing patient vitals and Electronic Medical Records while supporting video consultations.

    This layered approach optimizes resource allocation based on response time criticality and computational complexity, ensuring appropriate performance characteristics across diverse healthcare applications without unnecessary resource expenditure.

    This layered architecture directly addresses the availability and resilience requirements identified in Section~\ref{sec:threat-model}. By distributing processing across multiple tiers, the system maintains emergency response capabilities even if higher-layer Edge or Cloud nodes are compromised or unavailable. The hierarchical access control implemented at each layer constrains unauthorized access attempts by Type II and Type III adversaries, as compromise of a single layer does not grant access to all system resources.

    \subsection{Secure IoT-Cloud Healthcare Application Architecture}

\begin{figure}[htpb]
        \centerline{\includegraphics[width=93mm,height=45mm,scale=1]{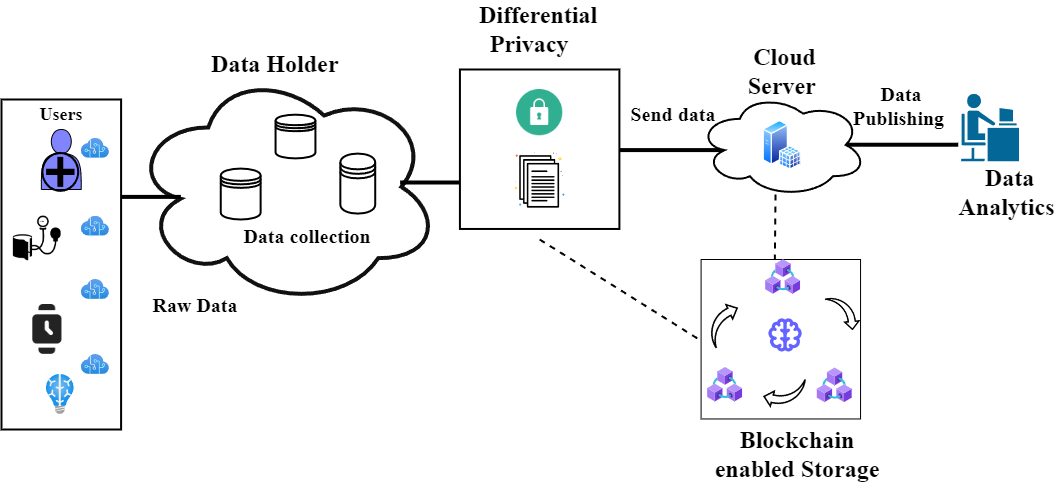}} 
        \caption{Securing Healthcare Data by DP and Blockchain} 
        \label{fig:smart-app-arch} 
    \end{figure}

    The proposed secure architecture for healthcare analytics integrates DP with distributed ledger technology to address the confidentiality, integrity, and accountability requirements established in Section~\ref{sec:threat-model}. Fig.~\ref{fig:smart-app-arch} illustrates the data flow and control mechanisms of this architecture. The security framework comprises six integrated components:\\ [6pt]
    \noindent
    \emph{(i) Data Collection:}  Patient data is aggregated from distributed IoT devices, wearables, and clinical systems. This heterogeneous data contains sensitive personally identifiable information (PII) requiring robust protection against the re-identification attacks described in Section~\ref{sec:threat-model}.\\ [6pt]
    \emph{(ii) Preprocessing:} Prior to privacy transformations, the data undergoes cleansing, normalization, and initial anonymization to remove direct identifiers and minimize re-identification risks through quasi-identifiers, addressing the linkage attack vectors identified in our threat model.\\ [6pt]
    \noindent
    \emph{(iii) Privacy Budget:} The framework implements a quantifiable privacy budget using the $\varepsilon$ parameter to constrain information leakage across multiple queries. This quantification establishes the maximum permissible privacy loss when adding or removing individual records from the dataset, directly addressing the composition attacks where adversaries accumulate information through repeated queries.\\ [6pt]
    \emph{(iv) Noise Injection:} Calibrated statistical noise is introduced to dataset calculations according to mathematically rigorous DP mechanisms. This process obscures individual data points while preserving statistical validity for aggregate analysis, providing formal guarantees against membership inference and attribute inference attacks by Type I adversaries.\\ [6pt]
    \emph{(v) Blockchain Security:} Blockchain technology is employed to address the integrity and accountability requirements from Section~\ref{sec:threat-model}, particularly the data tampering threats posed by Type II adversaries and audit trail tampering by Type III insider adversaries. The system utilizes off-chain storage mechanisms such as Ethereum off-chain solutions to mitigate transaction costs while maintaining data integrity across distributed stakeholders. Each transaction is cryptographically hashed and timestamped, creating tamper-evident records that detect any modification attempts. The current implementation employs the Raft consensus protocol, which provides crash fault tolerance (CFT) with deterministic finality, tolerating up to $f < n/2$ crashed nodes in a network of $n$ ordering service nodes. This CFT model operates under the assumption that ordering service operators within the permissioned consortium are trusted and do not exhibit arbitrary malicious behaviour. For deployment scenarios requiring resilience against Byzantine adversaries (Type II and Type III), the architecture is designed to accommodate a Byzantine fault-tolerant (BFT) ordering service, such as those supported by Hyperledger Fabric's pluggable consensus framework, which would provide integrity guarantees with up to $f < n/3$ arbitrarily compromised consensus nodes. Cryptographic transaction signing and hash chaining provide tamper-evidence at the ledger level independent of the consensus mechanism, ensuring that any post-consensus manipulation of committed blocks remains detectable.\\ [6pt]
    \emph{(vi) Privacy-Preserving Analytics:} The resulting differentially-private dataset enables sophisticated statistical and computational analyses without compromising individual privacy. This approach balances clinical utility with robust privacy guarantees, ensuring that even Type III insiders with elevated privileges cannot bypass DP protections, as noise injection occurs before data leaves the trusted computational environment.\\ [6pt]
    This dual-protection framework provides complementary defenses against the distinct threat classes in our adversary model: DP mechanisms protect against inference attacks on aggregate data and ML models (Type I and Type II threats), while distributed ledger technology ensures integrity verification and accountability through immutable audit trails (Type II and Type III threats). Hierarchical access control across computational layers constrains unauthorized access and limits the scope of compromise (all adversary types).

    \subsection{Procedure for Differential Privacy with Machine Learning}
\begin{figure}[htpb]
        \centerline{\includegraphics[width=0.5\linewidth]{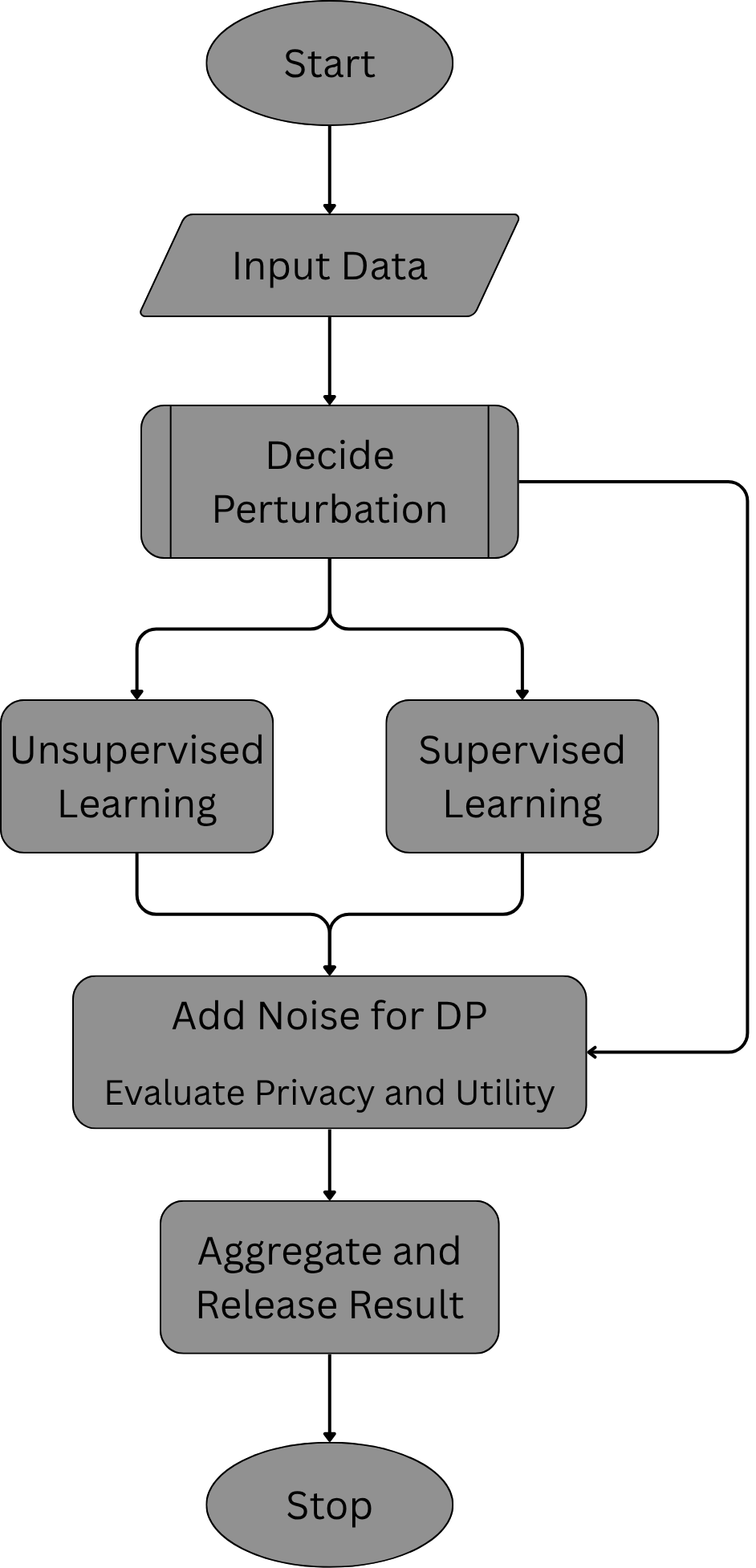}} 
        \caption{ML with DP Mechanism} 
        \label{fig:dp-mechanism} 
    \end{figure}

    The proposed methodology for implementing DP within ML workflows is illustrated in Fig.~\ref{fig:dp-mechanism}. This process comprises a structured sequence of operations designed to maintain analytical utility while providing mathematically rigorous privacy guarantees. \\ [6pt]
    \noindent
    \emph{(i) Data Acquisition:} The process begins with a dataset containing sensitive healthcare information that requires privacy protection while retaining analytical value.\\ [6pt]
    \emph{(ii) Data Preprocessing:} The dataset undergoes normalization, cleansing, and standardization procedures to ensure consistency and quality prior to analysis.\\ [6pt]
    \emph{(iii) ML Implementation:} Supervised or unsupervised learning algorithms are applied to the preprocessed data. For ensemble methods, this includes the generation of multiple decision trees for classification or regression tasks.\\ [6pt]
    \emph{(iv) Statistical Perturbation:} Controlled stochastic noise is introduced to the model outputs. This perturbation prevents adversarial reconstruction of individual data points while preserving aggregate statistical properties.\\ [6pt]
    \emph{(v) DP Mechanism:} A formal DP framework calibrates noise introduction according to sensitivity analysis and privacy budget constraints, ensuring mathematical guarantees of individual privacy.\\ [6pt]
    \emph{(vi) Aggregation Procedures:} Statistical measures such as cluster centroids, distribution parameters, and confidence intervals are calculated from the privacy-protected outputs, maintaining population-level insights while obscuring individual contributions.\\ [6pt]
    \emph{(vii) Analytical Interpretation:} The differentially private aggregated results are analyzed to extract clinically relevant insights, with careful attention to potential implications for healthcare decision-making.\\ [6pt]
    \emph{(viii) Privacy-Utility Assessment:} Quantitative evaluation of both privacy preservation (through DP guarantees) and analytical utility (through accuracy metrics) is performed to validate the methodology.\\ [6pt]
    \emph{(ix) Knowledge Synthesis:} The final privacy-protected analytical outcomes are synthesized into actionable healthcare intelligence that satisfies both privacy and utility requirements.\\ [6pt]
    \emph{(x) Process Completion:} The workflow concludes with documentation of privacy parameters and methodological constraints to ensure reproducibility and transparency.

    The algorithms detailed in the following subsections implement DP at various stages of ML workflows, providing formal guarantees against the gradient leakage and model inversion attacks identified in Section~\ref{sec:threat-model}. By introducing calibrated noise during training (input perturbation) or to model outputs (output perturbation), these mechanisms bound the influence of any single training example on model parameters, inherently providing robustness against model poisoning attacks by Type II adversaries.

    \subsection{Algorithms}
    This section delineates the integration of DP mechanisms with four ML algorithms: Random Forest, K-Means, Logistic Regression, and Naive Bayes. Each method incorporates Laplace, Gaussian, or hybrid noise calibrated to the sensitivity (\(\Delta\)) and privacy budget (\(\varepsilon\)), ensuring (\(\varepsilon,\delta\))-DP guarantees. Comprehensive pseudocode and procedural descriptions are provided for reproducibility and practical implementation.

    \subsubsection{Differentially Private Random Forest}
    The Random Forest ensemble is modified by injecting Gaussian noise into leaf node predictions post-training. Let \(f\) denote the sensitivity of the prediction function, computed as the maximum \(L_2\)-norm difference in outputs between adjacent datasets. For each leaf, Gaussian noise \(\mathcal{N}(0, \sigma^2)\) is sampled, where \(\sigma = \frac{\Delta f \sqrt{2\ln(1.25/\delta)}}{\varepsilon}\). The introduction of noise at the leaf level, rather than during the training process itself, represents a post-processing approach to DP that maintains the fundamental splitting criteria of the individual trees whilst privatising their outputs. Algorithm~\ref{algo:random-forest} formalises this process in detail.

    \begin{algorithm}
        \caption{Differentially Private Random Forest}
        \label{algo:random-forest}
        \KwData{Training data \(X\), labels \(y\), privacy budget \(\varepsilon\), failure probability \(\delta\), sensitivity \(\Delta\)}
        \KwResult{Differentially private random forest \(DP_{RF}\)}
        \SetKwFunction{AddNoise}{AddNoise}
        \SetKwFunction{SampleGaussian}{SampleGaussian}

        \ForEach{tree \(t \in\) RandomForest}{
            Train base decision tree: \(t \leftarrow \textsc{TrainTree}(X, y)\)\;
            \ForEach{leaf \(l \in t\)}{
                \(\sigma \leftarrow \Delta \sqrt{2\ln(1.25/\delta)} / \varepsilon\)\;
                \(noise \leftarrow \SampleGaussian(0, \sigma)\)\;
                \(l.\text{prediction} \leftarrow l.\text{prediction} + noise\)\;
            }
        }
        \KwRet{$DP_{RF}$}
    \end{algorithm}

    This algorithm ensures that each individual tree's predictions are perturbed with calibrated noise, whilst the overall ensemble maintains its predictive power through aggregation, as the independent noise additions tend to average out across multiple trees.

    \subsubsection{Differentially Private K-Means Clustering}
    The K-Means clustering algorithm, an unsupervised learning technique, requires modification to ensure DP during the iterative centroid refinement process. Cluster centroids are perturbed using Laplace noise during each iteration. Let \(\Delta f\) represent the \(L_1\)-sensitivity of centroid updates. At each iteration, Laplace noise \(\text{Lap}(\Delta f / \varepsilon)\) is added component-wise to each centroid coordinate.

    The iterative nature of K-Means presents a particular challenge for privacy preservation, as each update potentially leaks information. By calibrating the noise to the sensitivity and privacy budget at each step, the algorithm maintains its clustering efficacy whilst providing formal DP guarantees. Algorithm~\ref{algo:k-means} presents the complete procedure.

    \begin{algorithm}
        \caption{Differentially Private K-Means}
        \label{algo:k-means}
        \KwData{Dataset \(D\), clusters \(k\), max iterations \(T\), sensitivity \(\Delta\), \(\varepsilon\)}
        \KwResult{Private centroids \(C\), clusters \(S\)}
        \SetKwFunction{AssignClusters}{AssignClusters}
        \SetKwFunction{PerturbCentroid}{PerturbCentroid}

        \(C \leftarrow \textsc{InitialiseCentroids}(D, k)\)\;
        \For{\(t = 1\) \KwTo \(T\)}{
            \(S \leftarrow \AssignClusters(D, C)\)\;
            \ForEach{cluster \(c_i \in C\)}{
                \(\tilde{\mu}_i \leftarrow \text{mean}(S_i) + \text{Lap}(\Delta / \varepsilon)\)\;
                \(C \leftarrow C \cup \tilde{\mu}_i\)\;
            }
        }
        \KwRet{C, S}
    \end{algorithm}

    This differentially private adaptation of K-Means preserves the algorithm's ability to identify natural groupings in the data whilst ensuring that the resulting clusters and centroids do not compromise the privacy of individual observations in the dataset.

    \subsubsection{Differentially Private Logistic Regression}
    Logistic Regression, a cornerstone supervised learning algorithm for binary classification, can be rendered differentially private through gradient perturbation during the optimization process. This approach injects calibrated noise into the gradient computations used for parameter updates, ensuring that the trained model does not reveal sensitive information about individual training examples. Let \(\nabla L(w)\) denote the loss gradient with respect to model weights $w$. The sensitivity of this gradient, $\Delta$, is defined as the maximum \(L_1\)-norm difference in gradients that could result from adding or removing a single training example, calculated as \(\Delta = \max \|\nabla L_i(w)\|_1\). Laplace noise \(\eta \sim \text{Lap}(\Delta / \varepsilon)\) is injected into each gradient component prior to weight updates.

    This gradient perturbation approach represents an input perturbation method for DP, in contrast to the output perturbation used in Random Forest. By introducing noise during the training process itself, the algorithm ensures that the entire model fitting procedure maintains privacy guarantees. Algorithm~\ref{algo:logistic-regression} details this procedure.

    \begin{algorithm}
        \caption{Differentially Private Logistic Regression}
        \label{algo:logistic-regression}
        \KwData{Training set \(D\), \(\varepsilon\), \(\Delta\), learning rate \(\eta\), iterations \(T\)}
        \KwResult{Private weights \(w\)}
        \SetKwFunction{PerturbGradient}{PerturbGradient}

        Initialise \(w \leftarrow \mathbf{0}\)\;
        \For{\(t = 1\) \KwTo \(T\)}{
            Compute \(\nabla L(w) = \sum_{(x_i,y_i) \in D} (p_i - y_i)x_i\)\;
            \(\nabla L(w) \leftarrow \PerturbGradient(\nabla L(w), \Delta, \varepsilon)\)\;
            \(w \leftarrow w - \eta \nabla L(w)\)\;
        }
        \KwRet{w}
    \end{algorithm}

    The function \texttt{PerturbGradient} adds Laplace noise to each component of the gradient vector, with the noise magnitude calibrated according to the sensitivity and privacy budget. This approach ensures that the resulting logistic regression model maintains its predictive capabilities whilst providing formal privacy guarantees.

    \subsubsection{Differentially Private Naive Bayes}
    The Naïve Bayes classifier can be rendered differentially private through two complementary approaches depending on the feature representation, both presented in Algorithm~\ref{algo:naive-bayes}.

    \emph{Variant A (Discrete Features):} For datasets with discrete or binary features, the classical approach perturbs the sufficient statistics (counts) used to estimate conditional probabilities. For each feature \(x_j\) and class \(c_k\), Laplace noise drawn from \(\text{Lap}(1/\varepsilon)\) is added to both the feature-class count and the class count, where the sensitivity equals 1 since adding or removing a single record changes any individual count by at most one unit. Because independent noise on numerator and denominator does not guarantee that the resulting ratio lies in $[0,1]$ or that probabilities sum to unity across features, the perturbed estimates are clamped to $[0,1]$ and renormalized per class to form valid probability distributions.

    \emph{Variant B (Continuous Features):} For datasets with continuous features, input perturbation adds calibrated noise (Laplace, Gaussian, or hybrid) directly to the training feature values, after which a standard Gaussian Naive Bayes classifier estimates class-conditional mean and variance parameters from the perturbed data. This approach leverages the post-processing immunity property of DP, as the Gaussian likelihood estimation on already-privatised data does not consume additional privacy budget. Variant B is appropriate for healthcare datasets containing continuous clinical measurements (e.g., CD4 and CD8 counts) that are better suited to Gaussian likelihood estimation than discrete count models.

    \begin{algorithm}
        \caption{Differentially Private Naive Bayes}
        \label{algo:naive-bayes}
        \KwData{Training set \(D = \{(x_i, y_i)\}\), \(\varepsilon\), sensitivity \(\Delta\), classes \(C\), features \(F\)}
        \KwResult{Differentially private Naive Bayes model \(NB_{priv}\)}
        \SetKwFunction{PerturbLikelihood}{PerturbLikelihood}

        \tcp{Variant A: Discrete count perturbation}
        \ForEach{class \(c_k \in C\)}{
            \(n_k \leftarrow |\{x \in D : y = c_k\}|\)\;
            \ForEach{feature \(x_j \in F\)}{
                \(count_{j,k} \leftarrow \sum_{x \in D} \mathbf{1}(x_j=1 \land y=c_k)\)\;
                \(\tilde{P}(x_j|c_k) \leftarrow \frac{count_{j,k} + \text{Lap}(1/\varepsilon)}{n_k + \text{Lap}(1/\varepsilon)}\)\;
                \(\tilde{P}(x_j|c_k) \leftarrow \text{clamp}(\tilde{P}(x_j|c_k),\ 0,\ 1)\)\;
            }
            Renormalise: \(\tilde{P}(\cdot|c_k) \leftarrow \tilde{P}(\cdot|c_k) / \sum_j \tilde{P}(x_j|c_k)\)\;
        }
        \KwRet{\(\tilde{P}(x_j|c_k)\)}

        \BlankLine
        \tcp{Variant B: Continuous input perturbation}
        \ForEach{\(x_i \in D\)}{
            \(\tilde{x}_i \leftarrow x_i + \text{Noise}(\Delta, \varepsilon)\) \tcp{Laplace, Gaussian, or hybrid}
        }
        Fit Gaussian NB: \(\hat{\mu}_{j,k}, \hat{\sigma}^2_{j,k} \leftarrow \text{MLE}(\{\tilde{x}_i : y_i = c_k\})\)\;
        \KwRet{\(\hat{\mu}_{j,k}, \hat{\sigma}^2_{j,k}\ \forall\ j,k\)}
    \end{algorithm}

    \subsubsection{Theoretical Guarantees}
    Each of the described algorithms satisfies (\(\varepsilon,\delta\))-DP, with mathematical proofs derived from the fundamental properties of DP, particularly the post-processing immunity theorem. This theorem establishes that any function of a differentially private output remains differentially private, without requiring additional privacy budget expenditure.

    For the Random Forest algorithm, the addition of noise to leaf node predictions preserves privacy as predictions depend solely on these perturbed aggregates, with no further access to the original training data. The K-Means and Logistic Regression algorithms adhere to the sequential composition theorem, which quantifies the cumulative privacy loss across multiple operations on the same data. The Naïve Bayes discrete variant (Variant A) satisfies DP through the perturbation of sufficient statistics, with post-hoc clamping and renormalization preserving privacy guarantees via the post-processing immunity theorem. The continuous (Gaussian) variant (Variant B) inherits its privacy guarantee directly from the input perturbation step, as fitting class-conditional Gaussian distributions to already-privatised data constitutes post-processing of a differentially private output.

    These theoretical guarantees ensure that the privacy properties of the algorithms hold regardless of the adversary's computational power or background knowledge, providing a robust foundation for privacy-preserving ML in sensitive healthcare applications. The formal nature of these guarantees distinguishes DP from heuristic anonymization approaches that lack such mathematical rigor.

    \subsection{Blockchain-Based Trust Framework}

    To establish robust trust in our healthcare data ecosystem, we propose integrating blockchain technology with our multi-layered architecture and DP framework. This blockchain component creates a trusted foundation for healthcare transactions where centralized trust cannot be assumed, particularly in multi-stakeholder environments involving patients, providers, insurers, and researchers.

    \subsubsection{Trust Architecture using Blockchain}

    The proposed blockchain implementation addresses the fundamental trust challenges in healthcare data sharing by creating a decentralized trust framework where no single entity controls the entire system. This approach is particularly valuable in healthcare contexts, where patients must trust multiple parties with their sensitive information~\cite{Rangwala2025}.

    Our design employs a permissioned blockchain network where trusted healthcare entities (hospitals, clinics, research institutions, and regulatory agencies) serve as validator nodes. Unlike public blockchains, this consortium model balances efficiency with trusted verification processes \cite{Rachakonda2021}. The architecture establishes trust through: \\[6pt]
    \emph{(i) Distributed Consensus:} Critical healthcare transactions achieve validity only through agreement among multiple independent validators, eliminating single points of trust failure.\\ [6pt]
    \emph{(ii) Cryptographic Verification:} Digital signatures and hash functions create mathematical proof of data integrity, replacing institutional trust with cryptographic certainty.\\ [6pt]
    \emph{(iii) Immutable Record-Keeping:} The append-only structure of blockchain creates tamper-evident records, allowing stakeholders to trust the permanence and integrity of healthcare data histories.\\ [6pt]
    \emph{(iv) Trust Transparency:} The blockchain provides visibility into who accessed what data and when, creating accountability without requiring blind trust in any single record-keeper.

    \subsubsection{Establishing Trust Between Privacy and Utility}

    A significant innovation in our proposed framework is resolving the traditional trust tension between data privacy and research utility. By integrating blockchain with DP mechanisms, we create a system where:\\[6pt]
    (i) Patients can trust that their privacy is mathematically guaranteed through DP, with provable limits on information disclosure.\\ [6pt]
    (ii) Researchers can trust the authenticity and integrity of aggregated healthcare data without needing access to individual records.\\ [6pt]
    (iii) Regulators can trust the compliance status of the system through cryptographically verified audit trails.\\ [6pt]
    This balanced approach addresses the trust paradox in healthcare informatics, maintaining trust in both privacy protection and data utility simultaneously \cite{Jiang2022}. The blockchain would store cryptographic commitments to privacy budgets, creating verifiable records of compliance with privacy guarantees.

    \subsubsection{Smart Contracts as Trust Automation}

    The proposed framework would implement specialized smart contracts that codify trust relationships into executable agreements. These self-enforcing protocols would automate trust in consent by converting patient preferences into cryptographically enforced access rules, allowing patients to trust that their sharing preferences are honored without ongoing monitoring. They would create trustworthy audit trails by generating immutable records of every data access event that all stakeholders can independently verify, establishing trust through transparency. Additionally, they would enable trusted multi-party research by facilitating complex data sharing arrangements between competing institutions that might otherwise lack sufficient trust for collaboration.

    \subsubsection{Blockchain Architecture Specifications}

    The proposed blockchain integration employs a permissioned consortium architecture to address security and auditability requirements of privacy-preserving healthcare analytics. This subsection delineates the technical specifications for blockchain implementation within the multi-layer IoT-Cloud framework.\\ [6pt]
    \noindent
    \emph{(i) Platform Selection:} The architecture specifies Hyperledger Fabric~\cite{Androulaki2018} as the blockchain substrate, selected for its permissioned network access compatible with regulatory frameworks (HIPAA, GDPR), modular architecture enabling healthcare-specific customization, private data collections for confidential information sharing among authorized participants, and deterministic transaction finality through crash fault-tolerant ordering services. This selection was informed by established healthcare deployment precedents~\cite{Stamatellis2020} and enterprise-grade tooling.\\ [6pt]
    \emph{(ii) Consensus Mechanism:} The framework employs Raft consensus protocol~\cite{Ongaro2014}, providing crash fault tolerance with deterministic finality suitable for permissioned healthcare networks. The network topology comprises peer nodes deployed at the Edge layer maintaining ledger replicas and executing smart contract logic, ordering service nodes at Cloud infrastructure establishing transaction sequence and block generation, certificate authority infrastructure managing cryptographic identities and access credentials across the multi-layer architecture, and client applications at IoT and Edge layers submitting transactions through SDK interfaces.\\ [6pt]
    \emph{(iii) Smart Contract Architecture:} The blockchain layer implements chaincode (Hyperledger terminology for smart contracts) encoded in Go language. The Privacy Budget Ledger maintains immutable records of epsilon allocation, tracks cumulative privacy expenditure across analytical queries, enforces budget constraints through transaction validation logic, and records privacy parameters ($\varepsilon$, $\delta$, sensitivity $\Delta f$, noise distribution) for audit verification. The Data Provenance Chain establishes cryptographic linkage between data transformations across architectural layers, recording SHA-256 hashes at each processing stage (IoT, Edge and Cloud), maintaining temporal metadata such as ISO 8601 timestamps and processing duration, and enabling end-to-end traceability for regulatory compliance. The Access Control Enforcement implements attribute-based access control policies, logs access attempts with requestor identity verification, and records patient consent states for data processing operations.\\ [6pt]
    \emph{(iv) Transaction Structure:} Blockchain transactions encapsulate transaction identifier (UUID v4), timestamp with nanosecond precision, operation type enumeration, namely, \textsc{DataIngestion}, \textsc{DPQuery}, \textsc{BudgetUpdate}, and \textsc{AccessRequest}, cryptographic hash of data or query result (SHA-256 digest), privacy parameters encoded as JSON object containing $\{\varepsilon, \delta, \text{noiseType}, \Delta f\}$, digital signature (ECDSA with NIST P-256 curve) authenticating authorized entity, and metadata fields including device identifier, processing layer designation, and analytical purpose classification. This structure provides cryptographic proof of privacy guarantee compliance without exposing sensitive healthcare information.\\ [6pt]
    \emph{(v) Integration with Differential Privacy:} The blockchain interfaces with the DP module through REST API endpoints, enabling atomic operations that precede each analytical query. Prior to executing any DP-protected query, the system invokes the Privacy Budget Ledger smart contract to verify remaining budget sufficiency and record the impending privacy expenditure. This integration creates cryptographically verifiable proof that privacy budgets have not been exceeded, addressing the composition attack vulnerability where adversaries accumulate information through repeated queries~\cite{Dwork2014}. Post-query execution, the system commits query results as cryptographic hashes and privacy parameters to the ledger, establishing an immutable audit trail.\\ [6pt]
    \emph{(vi) Cryptographic Primitives:} The blockchain implementation employs standardized cryptographic algorithms. ECDSA with NIST P-256 curve (secp256r1) provides digital signatures with 128-bit security strength~\cite{Mehuron1996}. SHA-256 hash function ensures data integrity verification and block linking, conforming to NIST FIPS 180-4 specifications~\cite{NIST1993}. X.509 v3 certificates manage identity within the permissioned network, issued by the Hyperledger Fabric Certificate Authority. These selections align with NIST recommendations for cryptographic algorithm standards in healthcare information systems~\cite{NISTHealth}.\\ [6pt]
    This blockchain specification provides comprehensive technical grounding for implementing tamper-proof privacy guarantees within the proposed healthcare system. While this work focuses on architectural design and DP implementation validation through experimental evaluation (Section~\ref{sec:performance-eval}), the detailed blockchain specifications establish a rigorous foundation for future system integration.

    \subsubsection{Trust Recovery Mechanisms}

    A critical aspect of any trust framework is the ability to recover from failures. Our blockchain design incorporates mechanisms for fault detection, identifying crashed or unresponsive nodes through heartbeat monitoring and consensus timeouts. Under the current Raft-based CFT ordering service, detection is limited to crash faults; extending to Byzantine fault detection, capable of identifying arbitrarily malicious behaviour through consensus discrepancies, would require migration to a BFT ordering service. Trust revocation processes allow for removing compromised entities from trusted operations, while transparent remediation documents all trust violations and recovery actions on the blockchain itself \cite{Rangwala2025}.
    These mechanisms acknowledge that trust violations will occasionally occur while providing structured, transparent processes for maintaining system-wide trust even when individual components fail. The resilience of the trust framework is particularly important in healthcare contexts where continuity of operations directly impacts patient outcomes.

    \section{Implementation}
    \label{sec:implementation}
    The implementation of DP in healthcare systems necessitates meticulous handling of medical records, encompassing patient histories, diagnoses, and treatments, to balance privacy preservation with analytical utility. This section delineates the methodologies for noise injection, key considerations for medical data, and the framework for evaluating ML algorithms under DP constraints.

    \subsection{Noise addition in Differential Privacy}
    Noise introduction in DP ensures individual privacy while maintaining statistical validity. In healthcare contexts, noise is applied during aggregation (e.g., calculating disease prevalence) or statistical computations (e.g., average treatment duration). The calibration of noise magnitude depends on the sensitivity of the query and the privacy budget ($\varepsilon$), which governs the trade-off between privacy guarantees and data accuracy. Table~\ref{table:noise-comparison} presents a comparison of different noise mechanisms.\\

    \begin{enumerate} [(i) ]
    \item \emph{Laplace Noise Mechanism:} Laplace noise, drawn from a symmetric exponential distribution, is scaled by the sensitivity ($\Delta$) of the query and $\varepsilon$. This mechanism is optimal for low-dimensional datasets (e.g., patient counts per diagnosis) due to its heavy-tailed distribution, which provides robust privacy guarantees. However, excessive noise at low $\varepsilon$ values may degrade utility.\\

    \item \emph{Gaussian Noise Mechanism:} Gaussian noise, characterized by a bell-shaped normal distribution, is governed by $(\varepsilon, \delta)$-DP, where $\delta$ represents the probability of privacy leakage. It is suitable for high-dimensional data (e.g., electronic health records) due to its lighter tails, which preserve utility in complex analyses.\\

    \item \emph{Hybrid Laplace-Gaussian Noise Mechanism:} The proposed hybrid mechanism combines Laplace and Gaussian noise to leverage the complementary strengths of both distributions, Laplace's robust privacy guarantees through heavy tails and Gaussian's utility preservation through concentrated mass near zero. This approach addresses the limitation that Laplace noise may be excessive for high-dimensional data while Gaussian noise may provide insufficient protection for outlier-sensitive queries. Given a query function $f: \mathcal{D} \rightarrow \mathbb{R}^d$ with sensitivity $\Delta f$ and total privacy budget $\varepsilon_{\text{total}}$, the hybrid mechanism generates noisy output as:
    \begin{equation}
        M(D) = f(D) + \text{Lap}\left(\frac{\Delta f}{\varepsilon_L}\right) + \mathcal{N}\left(0, \frac{2\ln(1.25/\delta) \cdot \Delta f^2}{\varepsilon_G^2}\right)
    \end{equation}
    where $\varepsilon_L$ and $\varepsilon_G$ denote the privacy budget allocated to Laplace and Gaussian components respectively, satisfying $\varepsilon_L + \varepsilon_G = \varepsilon_{\text{total}}$ under sequential composition theorem~\cite{Dwork2014}, and $\delta$ is the probability of privacy failure for the Gaussian component. The privacy budget is partitioned between noise mechanisms based on data characteristics. For datasets with mixed dimensionality or varying sensitivity profiles, we employ an equal allocation strategy ($\varepsilon_L = \varepsilon_G = \varepsilon_{\text{total}}/2$) as the baseline. Alternative allocation strategies include Laplace-dominant ($\varepsilon_L = 0.7\varepsilon_{\text{total}}, \varepsilon_G = 0.3\varepsilon_{\text{total}}$) for low-dimensional data requiring stronger tail guarantees, and Gaussian-dominant ($\varepsilon_L = 0.3\varepsilon_{\text{total}}, \varepsilon_G = 0.7\varepsilon_{\text{total}}$) for high-dimensional data prioritizing utility preservation. By the sequential composition property of DP, applying two independent randomized mechanisms with privacy guarantees $\varepsilon_L$-DP and $\varepsilon_G$-DP respectively yields an overall privacy guarantee of $(\varepsilon_L + \varepsilon_G)$-DP = $\varepsilon_{\text{total}}$-DP~\cite{Dwork2014}. The resulting hybrid noise distribution exhibits intermediate tail behavior; heavier than pure Gaussian but lighter than pure Laplace, creating a more nuanced privacy-utility trade-off. This distribution decays gradually over a larger range than either individual distribution, making it particularly suitable for protecting datasets with mixed dimensionality or varying sensitivity profiles. The moderate tail behavior provides robust privacy guarantees similar to Laplace while preserving analytical utility comparable to Gaussian noise, especially beneficial for healthcare datasets containing both low-dimensional aggregates (patient counts) and high-dimensional features (electronic health records). The optimal allocation of the privacy budget between components can be determined through empirical evaluation via grid search over the allocation parameter $\alpha \in [0,1]$ where $\varepsilon_L = \alpha \varepsilon_{\text{total}}$ and $\varepsilon_G = (1-\alpha)\varepsilon_{\text{total}}$.\\
    \end{enumerate}

    \begin{table*}[htbp]
        \centering
        \caption{Comparison of Noise}
        \label{table:noise-comparison}
        \scriptsize
        \begin{tabular}{|p{3cm}|p{4cm}|p{4cm}|p{4cm}|}
            \hline

            \textbf{Factor} & \textbf{Laplace Noise} & \textbf{Gaussian Noise} & \textbf{Combined L-G} \\
            \hline
            Probability Distribution & Laplace distribution & Gaussian distribution & Combined distribution \\
            \hline
            Symmetry & Symmetric around 0 & Symmetric around 0 & Symmetric around 0 \\
            \hline
            Scale & Controlled by sensitivity & Controlled by sensitivity & Controlled by sensitivity \\
            \hline
            Privacy Budget & $\varepsilon$ parameter & $\varepsilon$ parameter & $\varepsilon$ parameter \\
            \hline
            Tail Behavior & Heavier tails & Lighter tails & Moderate tails \\
            \hline
            Data Type & Suitable for bounded data, lower dimension & Suitable for any data type, higher dimension & Suitable for any data type, higher dimension \\
            \hline
            Adding Noise Mechanism & Add noise to each data point & Add noise to aggregate & Add to individual or aggregate\\
            \hline
            Trade-off & More noise, high privacy & Lesser noise, high privacy & Moderate noise, high privacy \\
            \hline

        \end{tabular}
    \end{table*}

    \subsection{Key Considerations in Differential Privacy for Medical Health Data}
    \begin{enumerate} [(i) ]
    \item \emph{Privacy Budget}: Cumulative $\varepsilon$ across multiple queries must be constrained to prevent excessive privacy loss. Dynamic budgeting strategies, such as zero-concentrated DP, can optimize allocations for longitudinal studies.

    \item \emph{Sensitivity of Data}: Attributes like genetic markers or rare diseases necessitate higher noise due to their identifiability risks. Sensitivity analysis should precede noise calibration.

    \item \emph{Noise Mechanism Selection}: Laplace suits bounded, low-dimensional queries (e.g., disease counts), while Gaussian is preferable for unbounded, high-dimensional analyses (e.g., ML model training).

    \item \emph{Granularity vs Utility}: Aggregating data (e.g., age groups instead of exact ages) reduces sensitivity but may obscure critical patterns. Context-aware aggregation preserves utility for clinical decision-making.

    \item \emph{Analysis Type}: The type of analysis being performed matters. Simple counts or averages may require less noise compared to complex statistical analyses or ML tasks.

    \item \emph{Data Size}: Larger data sets can tolerate more noise, while smaller data sets might need careful handling to avoid excessive distortion.

    \item \emph{Privacy vs Utility}: A critical consideration is the trade-off between preserving individual privacy and maintaining the usefulness of the data. Striking the right balance is essential to ensure meaningful results without compromising privacy.

    \item \emph{Regulations and Standards}: HIPAA and GDPR mandate strict anonymization. DP parameters must align with legal thresholds for de-identification.

    \item \emph{Expert Consultation}: Input from clinicians ensures noise levels do not invalidate medical insights, while privacy experts validate compliance with ethical standards.
    \end{enumerate}


    \subsection{Evaluation of ML Algorithms with DP techniques}
    Healthcare research and analytics require accurate data for delivering correct and error free decisions beneficial to humanity. Healthcare researchers and analysts utilise medical datasets to improve community health indices, hospital systems and public health policies. Various analytical approaches are required for different situations, including: (i) descriptive (e.g., number of hospitalised patients in the previous week), (ii) diagnostic (e.g., hospitalization causes), (iii) predictive (e.g., likely hospitalizations in the coming week) and (iv) prescriptive (e.g., preventative medicine recommendations).
    Consequently, different ML algorithms are employed in healthcare analytics. For instance, patients visiting hospitals may be grouped according to symptom intensity using the K Means ML algorithm to cluster patients into k=3 groups (no symptoms/mild symptoms/strong symptoms). K means is an unsupervised ML algorithm that clusters data points into groups based on similarity. To predict infection likelihood, Logistic Regression is valuable, identifying relationships between patient features such as comorbidity, age and present symptoms to generate binary predictions about future infection status. Such predictions facilitate hospital/bed/medicine capacity planning and preventative care provision. Naive Bayes, a supervised ML algorithm using labelled data, classifies instances into predefined classes based on independent features, aiding diagnosis. For example, Naive Bayes can identify jaundice when symptoms include yellow eye colouration, turbid urine and elevated body temperature. Random Forest ML algorithms are employed for predicting drug sensitivity.
    DP is applied to datasets to enable ML analysis whilst protecting individual privacy. Generally, K means, Logistic Regression, Random Forest and Naive Bayes cannot be directly compared as they address different tasks. A primary objective of this work is to assess ML query/analytics accuracy on differentially private datasets compared to accuracy on original datasets. Therefore, experiments evaluate accuracy against privacy budget $\varepsilon$.
    The efficacy of DP lies in its ability to maintain data usability whilst protecting individual privacy, measured by accuracy:
    \[
        \text{Accuracy} = \frac{TP + TN}{TP + FP + TN + FN}
    \]
    \parskip 0pt
    Where,\\
    \emph{TP (True Positive)} i.e. the count of positive outcomes correctly classified under positive class \\
    \emph{TN (True Negative)} i.e. the count of negative outcomes correctly classified under negative class \\
    \emph{FP (False Positive)} i.e. the count of negative outcomes incorrectly classified under positive class \\
    \emph{FN (False Negative)} i.e. the count of positive outcomes that are incorrectly classified under negative class \\ [6pt]




    \section{Performance Evaluation}
    \label{sec:performance-eval}
    This section presents the experimental methodology, dataset characteristics, and systematic assessment of DP techniques. The efficacy of the proposed differentially private data aggregation methods, each employing different noise characteristics for various ML models, is evaluated and analyzed\footnote{Code and experiment artifacts are available at: \url{https://github.com/Cloudslab/DP-Healthcare}}.

    \subsection{Experimental Setting}

    Experiments were conducted in Python 3.9.6 using scikit-learn for ML algorithms, NumPy for numerical computation, and pandas for data processing. Differential privacy mechanisms were implemented via custom functions to control noise calibration and privacy budget allocation. All experiments used fixed random seeds for reproducibility. Security validation employed a publicly available medical dataset from the UCI repository, retrieved programmatically using the \texttt{ucimlrepo} package, a standard benchmark for privacy-preserving healthcare ML.

    Multi-layer architecture performance was evaluated through discrete-event simulations incorporating stochastic models of network and computational behavior. Network delays were modeled as uniformly distributed variables calibrated against large-scale RTT measurements reported by Charyyev et al.~\cite{Charyyev2020}, who measured ping latency from 8,456 end-users to 6,341 edge servers and 69 cloud locations. Their results show that 58\% of users reach a nearby edge server in under 10~ms, with median edge RTT of approximately 5--10~ms for the majority of users, while cloud RTT typically ranges from 30--100~ms. Our simulation parameters---IoT-to-Edge (2--8~ms) and Edge-to-Cloud (40--80~ms)---fall within these empirically observed ranges, representing network-layer round-trip time for hospital-proximate edge deployments and well-provisioned WAN connections respectively. Data transfer latency was computed deterministically from payload size and bandwidth assumptions. Computational processing times, including ML inference were modeled as uniform distributions reflecting hardware capabilities at each tier. Blockchain consensus performance was simulated using Hyperledger Fabric’s Raft protocol across varying topologies (4–13 nodes) to assess integrity verification overhead. Monte Carlo evaluation with 100 independent trials per configuration provided statistical confidence intervals for all performance metrics.

    \subsection{Data Set}
    The evaluation employed the UCI AIDS Clinical Trials Group Study 175 dataset, which comprises healthcare statistics from AIDS patients \cite{Hammer1996}. This dataset encompasses both medical parameters (CD4 counts at baseline and months 20, CD8 counts at baseline and month 20, prescribed medications) and demographic attributes (date, time, age, ethnicity, and gender) that constitute sensitive information whose disclosure could potentially result in privacy violations and social implications. The UCI AIDS dataset was selected for its appropriate dimensionality, containing 2,139 observations across 27 variables, of which 23 serve as input features for model training ($d = 23$). The dataset exhibits significant class imbalance with 75.7\% of samples in class 0 (censored) and 24.3\% in class 1 (failure), reflecting realistic clinical trial outcomes. To address this imbalance, we employed undersampling to create balanced training sets with equal representation (417 samples per class, 834 total), improving model convergence and preventing majority-class bias under DP noise. Table~\ref{table:dataset-composition} reports the exact instance counts per class at each stage of the data pipeline to facilitate independent replication. The dataset was partitioned into 80\% training and 20\% test sets using stratified random splitting (random seed 42) to preserve class proportions. No separate validation set was employed; hyperparameter selection was based on established defaults from the literature rather than data-driven tuning, ensuring that the full training set was available for DP noise calibration.

    \begin{table}[!ht]
        \centering
        \scriptsize
        \caption{Dataset Composition: Instance Counts per Class at Each Pipeline Stage}
        \label{table:dataset-composition}
        \resizebox{\columnwidth}{!}{\begin{tabular}{|l|r|r|r|}
            \hline
            \textbf{Stage} & \textbf{Class 0} & \textbf{Class 1} & \textbf{Total} \\
            \hline
            Full dataset & 1618 (75.6\%) & 521 (24.4\%) & 2139 \\
            \hline
            Train (80\%, stratified) & 1294 (75.6\%) & 417 (24.4\%) & 1711 \\
            Test (20\%, stratified) & 324 (75.7\%) & 104 (24.3\%) & 428 \\
            \hline
            Train after undersampling & 417 (50.0\%) & 417 (50.0\%) & 834 \\
            Discarded (majority class) & 877 & --- & 877 \\
            \hline
        \end{tabular}}
    \end{table}

    \subsection{Experiment}
    The dataset was processed using K-Means clustering, Logistic Regression, Random Forest, and Naive Bayes algorithms with three DP techniques applied: Laplace, Gaussian, and hybrid Laplace-Gaussian noise mechanisms. All combinations of ML algorithms and DP techniques were evaluated across varying epsilon values ($\varepsilon \in \{0.5, 1.0, 2.0, 3.0, 5.0, 10.0\}$) to quantify accuracy degradation following DP obfuscation and identify practical epsilon thresholds for healthcare analytics. This methodical approach determined the optimal noise distribution and epsilon configuration for protecting ML training datasets while preserving analytical utility.

    The privacy budget parameters were extended to include higher epsilon values to investigate the practical epsilon threshold where input perturbation becomes viable for healthcare applications. The range spans from $\varepsilon = 0.5$ (strong privacy for highly sensitive HIV/AIDS patient data, aligning with recommendations for stigmatizing medical conditions~\cite{Dankar2013}) to $\varepsilon = 10.0$ (moderate privacy, consistent with real-world deployments such as the U.S. Census Bureau~\cite{Abowd2018}). This extended spectrum enables evaluation across different healthcare use cases: individual patient records (lower $\varepsilon$) to population-level analytics (higher $\varepsilon$), balancing re-identification risk with model utility as mandated by healthcare privacy regulations.

    The DP implementation employed input perturbation, where calibrated noise is added to training data before model learning. This approach ensures privacy guarantees independent of the learning algorithm. Sensitivity was established through per-record norm clipping, a standard technique for bounding query sensitivity in DP~\cite{Dwork2014}. Specifically, each record $x_i$ in the standardised AIDS dataset was projected onto the $L_2$ ball of radius $C$ via the clipping operation $\bar{x}_i = x_i \cdot \min(1, C / \|x_i\|_2)$, ensuring that $\|\bar{x}_i\|_2 \leq C$ for all records. The clipping threshold was set to the 95th percentile of the empirical per-record $L_2$ norm distribution ($C \approx 6.47$), a heuristic that balances bounded sensitivity against data distortion: lower thresholds increase clipping-induced bias while higher thresholds necessitate larger noise magnitudes. We note that the threshold selection itself is data-dependent; in a deployment setting, $C$ should be determined from public domain knowledge or a held-out sample with separate privacy accounting.

    $L_2$ clipping directly establishes the $L_2$ sensitivity $\Delta_2 = C$ required by the Gaussian mechanism. For the Laplace mechanism, which requires $L_1$ sensitivity, the Cauchy-Schwarz inequality yields the bound $\Delta_1 = \max\|\bar{x}_i\|_1 \leq C\sqrt{d}$, where $d$ is the number of features used in model training ($d = 23$ for the standardised AIDS dataset). In our experimental implementation, Laplace noise was calibrated to the $L_2$ bound ($\lambda = C/\varepsilon$) rather than the $L_1$ bound ($\lambda = C\sqrt{d}/\varepsilon$). This calibration provides a relaxed privacy guarantee for the Laplace mechanism: the effective privacy parameter is $\varepsilon_{\text{eff}} = \varepsilon\sqrt{d}$ $\approx 4.8\varepsilon$ rather than the nominal $\varepsilon$. Consequently, the Laplace results in Tables~\ref{table:epsilon-vs-accuracy} and~\ref{table:comprehensive-metrics} should be interpreted at this effective privacy level (e.g., nominal $\varepsilon = 1.0$ corresponds to $\varepsilon_{\text{eff}} \approx 4.8$ for Laplace). The Gaussian mechanism results carry the stated $(\varepsilon, \delta)$-DP guarantees exactly, and the hybrid mechanism inherits its guarantee from its component budgets under sequential composition. Relative accuracy comparisons between noise mechanisms at matched nominal $\varepsilon$ remain valid for algorithm selection guidance, as all mechanisms received noise calibrated to the same $L_2$ bound. In our dataset, approximately 5\% of records underwent clipping, introducing bounded distortion dominated by the subsequent DP noise at all tested privacy budgets. Gaussian noise standard deviation was set to $\sigma = C\sqrt{2\ln(1.25/\delta)}/\varepsilon$ with $\delta = 10^{-5}$, and the hybrid mechanism employed a weighted-noise combination: $\text{noise} = \alpha \cdot \text{Laplace}(C/\varepsilon) + (1-\alpha) \cdot \text{Gaussian}(\sigma)$ with $\alpha = 0.5$, ensuring distributional consistency across all data points and preserving algorithm-specific statistical assumptions~\cite{Dwork2014}.

    The DP algorithm was executed with five independent runs to ensure statistical robustness, with mean performance reported. K-Means utilized two clusters (matching the binary classification task) determined through optimal cluster-to-class mapping using the Hungarian algorithm~\cite{Kuhn1955}. Random Forest employed a five-level decision tree architecture with 100 trees, minimum 20 samples per split, and 10 samples per leaf to prevent overfitting. Logistic Regression used L2 regularization ($C=1.0$) with maximum 1000 iterations for convergence. Naive Bayes employed Variant B of Algorithm~\ref{algo:naive-bayes} (continuous input perturbation with Gaussian likelihood estimation) with default priors. Evaluation metrics include accuracy, precision, recall, F1-score, AUC, computed on a held-out test set (20\% of data, 428 samples) using stratified splitting to preserve class distribution, as detailed in Tables~\ref{table:baseline-performance},~\ref{table:epsilon-vs-accuracy},~\ref{table:comprehensive-metrics}, and~\ref{table:hybrid-alpha-analysis}.

    \subsection{Result Analysis}

\begin{figure}[!ht]
        \centering
        \includegraphics[width=\linewidth]{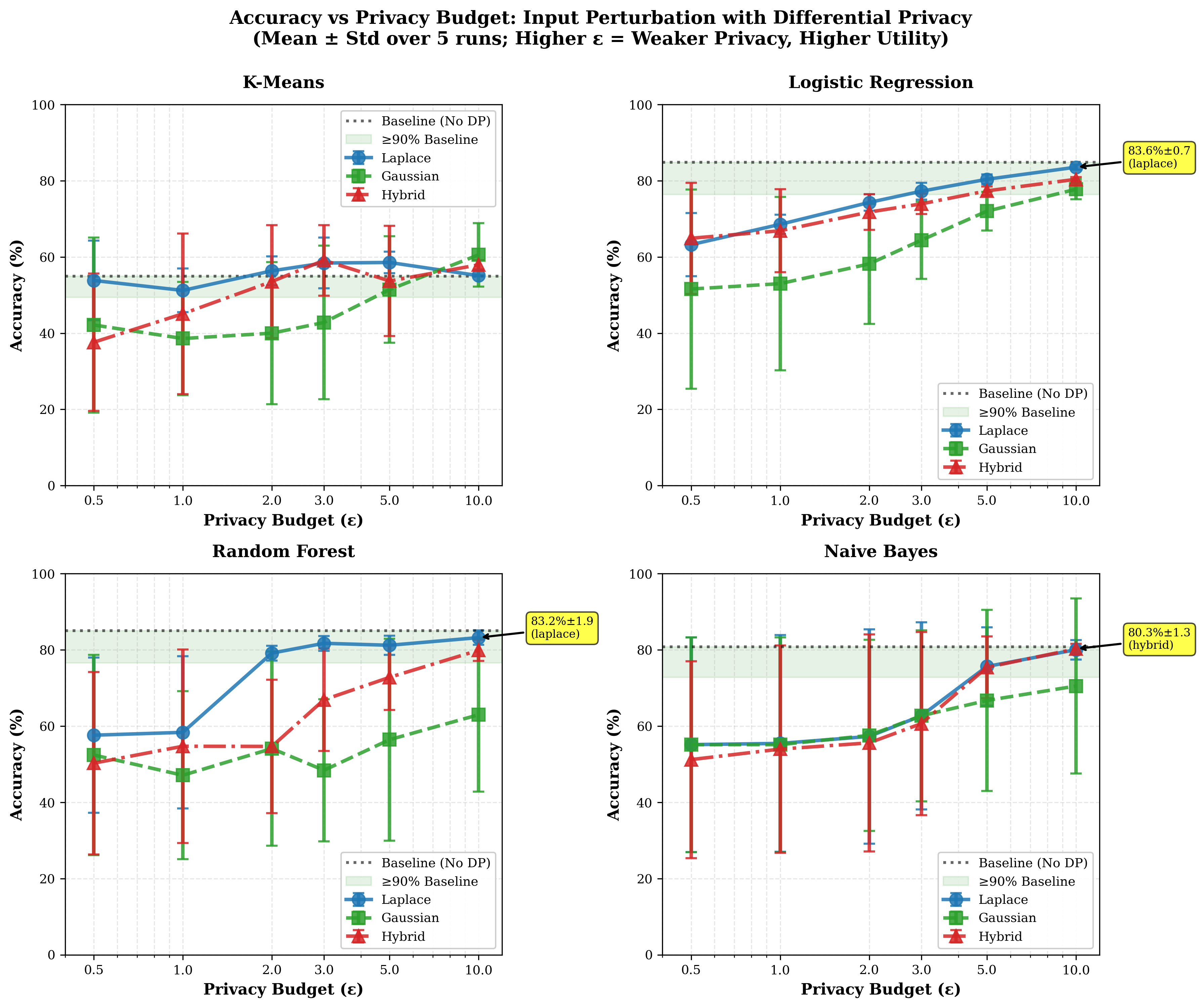}
        \caption{Accuracy of ML algorithms across different noise mechanisms and varying privacy budgets. Horizontal dotted lines indicate baseline performance without DP. Green shaded regions show $\geq$90\% baseline retention zone. Logarithmic x-axis emphasizes the wide privacy budget range tested.}
        \label{fig:dp-accuracy-comparison}
    \end{figure}

\begin{figure}[!ht]
        \centering
        \includegraphics[width=\linewidth]{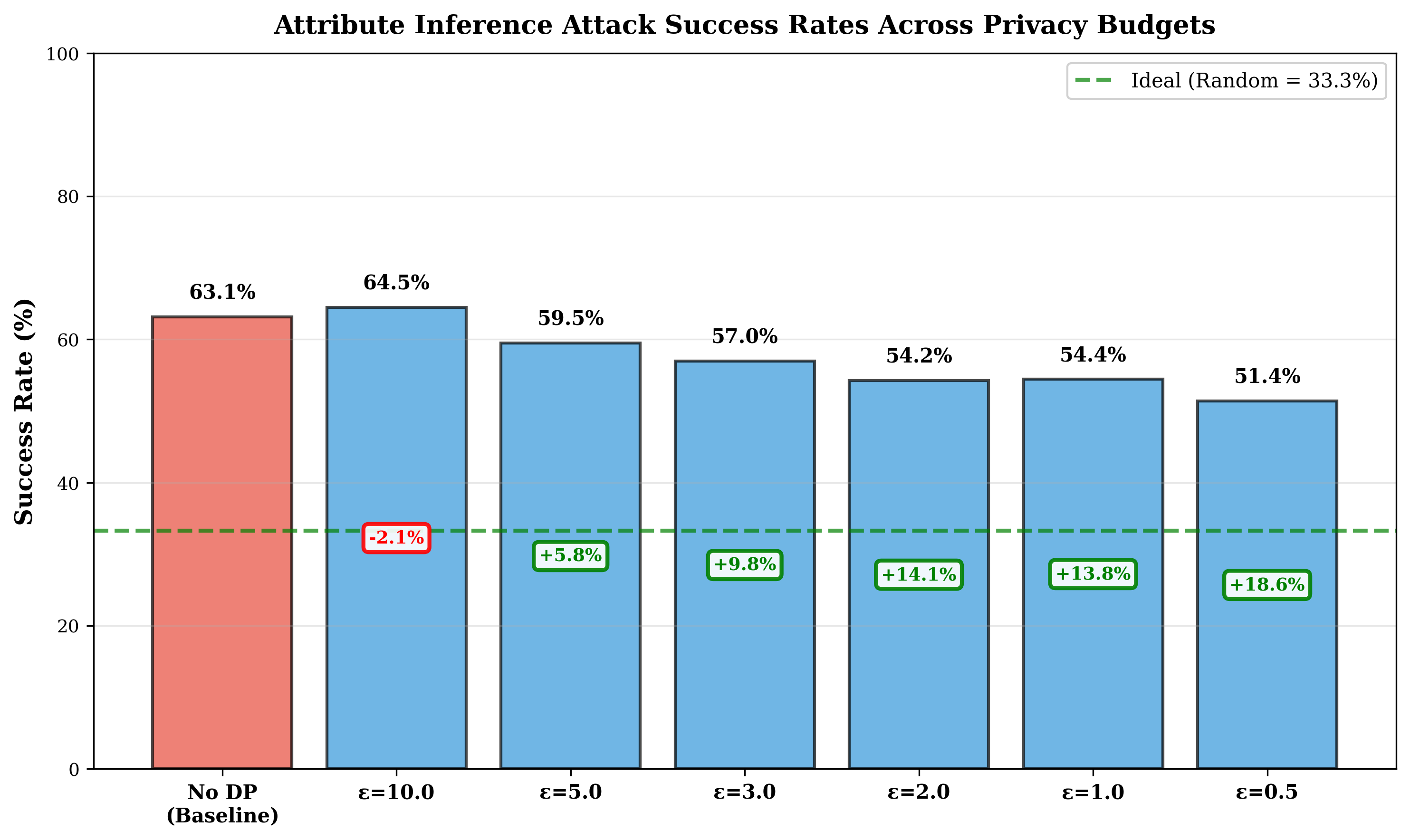}
        \caption{Attribute inference attack success rates across privacy budgets. Results averaged across all algorithms. Lower values indicate better privacy protection, with 33.3\% representing random guessing for the three-class sensitive attribute (ideal privacy).}
        \label{fig:attribute-inference-attack}
    \end{figure}

\begin{figure}[!ht]
        \centering
        \includegraphics[width=\linewidth]{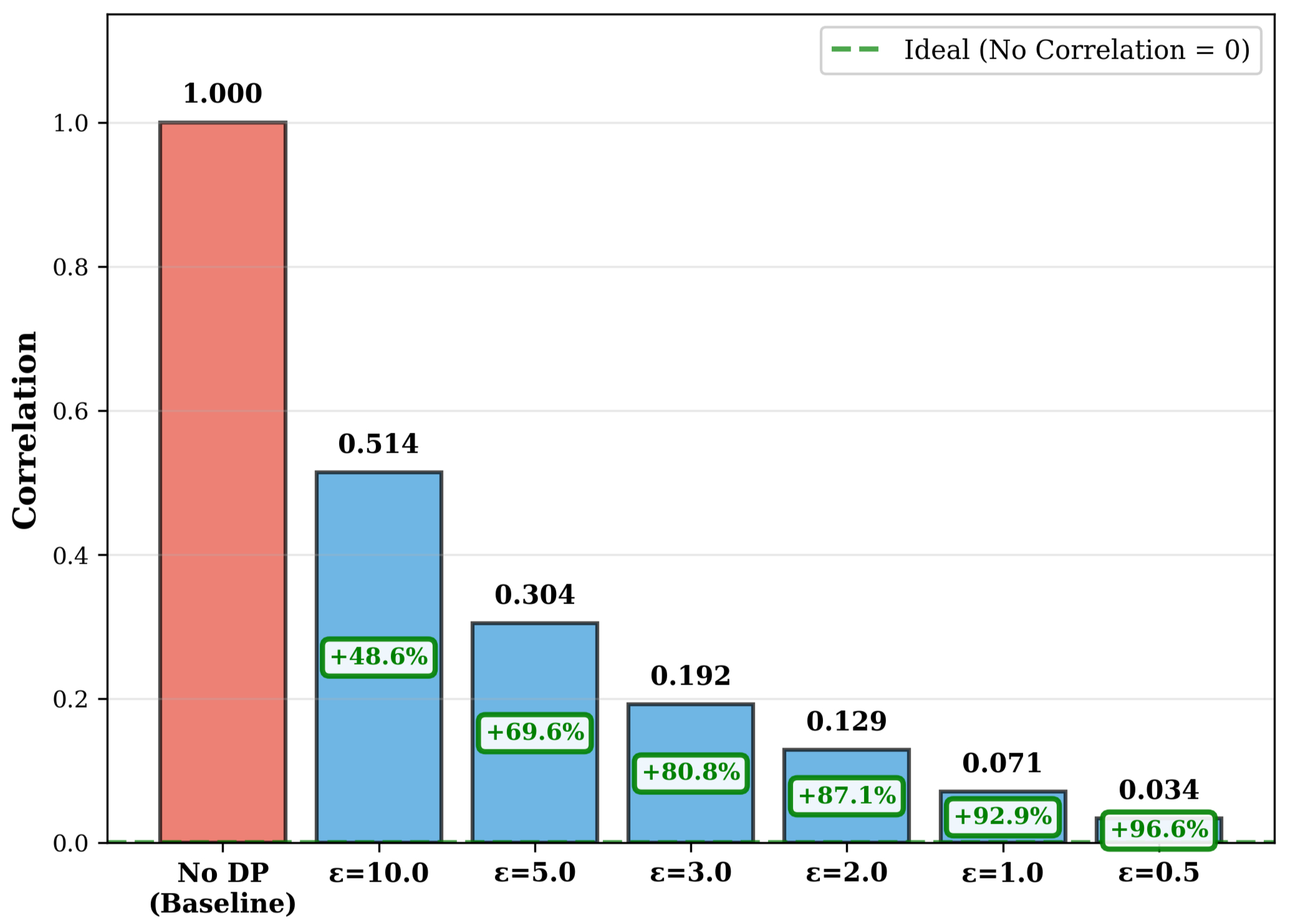}
        \caption{Data reconstruction attack correlation across privacy budgets. Results averaged across all algorithms. Lower correlation values indicate better privacy protection, with 0 representing no reconstruction capability (ideal privacy). The baseline bar is clipped with hatching to improve readability.}
        \label{fig:reconstruction-attack}
    \end{figure}

    Table~\ref{table:baseline-performance} presents baseline performance without DP, establishing reference metrics for privacy-utility trade-off analysis. Table~\ref{table:epsilon-vs-accuracy} presents the accuracy of various supervised and unsupervised ML techniques across an extended epsilon range ($\varepsilon \in \{0.5, 1.0, 2.0, 3.0, 5.0, 10.0\}$) using Laplace, Gaussian, and weighted-combination hybrid noise distributions. Table~\ref{table:comprehensive-metrics} provides detailed precision, recall, F1-score, and AUC metrics for $\varepsilon = 10.0$, demonstrating near-baseline performance recovery at moderate privacy levels. These values are plotted graphically in Fig.~\ref{fig:dp-accuracy-comparison}, illustrating the accuracy of DP-enabled K-Means, Logistic Regression, Random Forest and Naive Bayes respectively. Additionally, Figs.~\ref{fig:attribute-inference-attack} and~\ref{fig:reconstruction-attack} quantify privacy protection through adversarial evaluation, measuring attribute inference and data reconstruction attack success rates across privacy budgets.

    \begin{table}[!ht]
        \centering
        \scriptsize
        \caption{Baseline Model Performance (No Differential Privacy)}
        \label{table:baseline-performance}
        \begin{tabular}{|l|c|c|c|c|c|}
            \hline
            \textbf{Algorithm} & \textbf{Acc.} & \textbf{Prec.} & \textbf{Recall} & \textbf{F1} & \textbf{AUC} \\
            \hline
            Logistic Regression & 84.8 & 0.66 & 0.79 & 0.72 & 0.887 \\
            Random Forest & 85.0 & 0.64 & 0.88 & 0.74 & 0.922 \\
            Naive Bayes & 80.8 & 0.58 & 0.77 & 0.66 & 0.844 \\
            K-Means & 54.9 & 0.31 & 0.70 & 0.43 & 0.650 \\
            \hline
        \end{tabular}
    \end{table}

    \begin{table}[!ht]
        \centering
        \scriptsize
        \caption{$\varepsilon$ vs. Accuracy (\%): Mean$\pm$Std over 5 Runs (Hybrid: $\alpha=0.5$)}
        \label{table:epsilon-vs-accuracy}
        \begin{tabular}{|c|c|c|c|}
            \hline
            \textbf{Noise Type/} & & & \\
            \textbf{$\varepsilon$ Value} &  \textbf{Laplace} & \textbf{Gaussian} & \textbf{Hybrid} \\
            \hline
            \multicolumn{4}{|c|}{\textbf{K-Means}} \\
            \hline
            0.5 & 53.8$\pm$10.5 & 42.1$\pm$23.0 & 37.6$\pm$18.0 \\
            1.0 & 51.2$\pm$5.7 & 38.6$\pm$14.8 & 45.1$\pm$21.1 \\
            2.0 & 56.4$\pm$3.8 & 40.0$\pm$18.6 & 53.6$\pm$14.8 \\
            3.0 & 58.4$\pm$6.6 & 42.8$\pm$20.1 & 59.1$\pm$9.2 \\
            5.0 & 58.6$\pm$2.8 & 51.4$\pm$14.0 & 53.7$\pm$14.5 \\
            10.0 & 55.1$\pm$0.1 & 60.6$\pm$8.3 & 57.9$\pm$1.2 \\
            \hline
            \multicolumn{4}{|c|}{\textbf{Logistic Regression}} \\
            \hline
            0.5 & 63.2$\pm$8.3 & 51.6$\pm$26.1 & 64.9$\pm$14.6 \\
            1.0 & 68.6$\pm$2.5 & 53.0$\pm$22.7 & 66.9$\pm$10.9 \\
            2.0 & 74.3$\pm$2.2 & 58.2$\pm$15.8 & 71.8$\pm$4.7 \\
            3.0 & 77.2$\pm$2.2 & 64.4$\pm$10.1 & 73.9$\pm$2.7 \\
            5.0 & 80.4$\pm$1.3 & 72.0$\pm$5.0 & 77.3$\pm$1.1 \\
            10.0 & 83.6$\pm$0.7 & 77.8$\pm$2.7 & 80.4$\pm$0.6 \\
            \hline
            \multicolumn{4}{|c|}{\textbf{Random Forest}} \\
            \hline
            0.5 & 57.6$\pm$20.3 & 52.4$\pm$26.3 & 50.3$\pm$23.9 \\
            1.0 & 58.4$\pm$20.0 & 47.1$\pm$22.0 & 54.7$\pm$25.4 \\
            2.0 & 79.2$\pm$1.9 & 54.1$\pm$25.5 & 54.7$\pm$17.5 \\
            3.0 & 81.7$\pm$1.9 & 48.4$\pm$18.7 & 66.9$\pm$13.4 \\
            5.0 & 81.2$\pm$2.5 & 56.4$\pm$26.5 & 72.8$\pm$8.6 \\
            10.0 & 83.2$\pm$1.9 & 63.0$\pm$20.2 & 79.9$\pm$2.8 \\
            \hline
            \multicolumn{4}{|c|}{\textbf{Naive Bayes}} \\
            \hline
            0.5 & 55.1$\pm$28.2 & 55.1$\pm$28.2 & 51.2$\pm$25.8 \\
            1.0 & 55.5$\pm$28.5 & 55.1$\pm$28.2 & 54.0$\pm$27.2 \\
            2.0 & 57.3$\pm$28.1 & 57.6$\pm$25.1 & 55.6$\pm$28.5 \\
            3.0 & 62.7$\pm$24.5 & 62.7$\pm$22.5 & 60.7$\pm$24.0 \\
            5.0 & 75.7$\pm$10.3 & 66.7$\pm$23.8 & 75.4$\pm$8.1 \\
            10.0 & 80.0$\pm$2.6 & 70.5$\pm$23.0 & 80.3$\pm$1.3 \\
            \hline
        \end{tabular}
    \end{table}

    \begin{table}[!ht]
        \centering
        \scriptsize
        \caption{Comprehensive Performance Metrics at $\varepsilon = 10.0$: Mean$\pm$Std over 5 Runs (Hybrid: $\alpha=0.5$)}
        \label{table:comprehensive-metrics}
        \begin{tabular}{|l|c|c|c|c|}
            \hline
            \textbf{Algorithm/Noise} & \textbf{Acc.} & \textbf{Prec.} & \textbf{Recall} & \textbf{F1} \\
            \hline
            \multicolumn{5}{|c|}{\textbf{Logistic Regression}} \\
            \hline
            Laplace & 83.6$\pm$0.7 & 0.63$\pm$0.02 & 0.79$\pm$0.02 & 0.70$\pm$0.01 \\
            Gaussian & 77.8$\pm$2.7 & 0.53$\pm$0.04 & 0.74$\pm$0.03 & 0.62$\pm$0.03 \\
            Hybrid & 80.4$\pm$0.6 & 0.57$\pm$0.01 & 0.77$\pm$0.04 & 0.65$\pm$0.01 \\
            \hline
            \multicolumn{5}{|c|}{\textbf{Random Forest}} \\
            \hline
            Laplace & 83.2$\pm$1.9 & 0.63$\pm$0.04 & 0.78$\pm$0.02 & 0.69$\pm$0.03 \\
            Gaussian & 63.0$\pm$20.2 & 0.43$\pm$0.15 & 0.71$\pm$0.17 & 0.50$\pm$0.08 \\
            Hybrid & 79.9$\pm$2.8 & 0.58$\pm$0.05 & 0.71$\pm$0.10 & 0.63$\pm$0.05 \\
            \hline
            \multicolumn{5}{|c|}{\textbf{Naive Bayes}} \\
            \hline
            Laplace & 80.0$\pm$2.6 & 0.57$\pm$0.04 & 0.74$\pm$0.05 & 0.64$\pm$0.02 \\
            Gaussian & 70.5$\pm$23.0 & 0.56$\pm$0.18 & 0.63$\pm$0.25 & 0.53$\pm$0.11 \\
            Hybrid & 80.3$\pm$1.3 & 0.58$\pm$0.03 & 0.69$\pm$0.03 & 0.63$\pm$0.01 \\
            \hline
            \multicolumn{5}{|c|}{\textbf{K-Means}} \\
            \hline
            Laplace & 55.1$\pm$0.1 & 0.31$\pm$0.00 & 0.70$\pm$0.00 & 0.43$\pm$0.00 \\
            Gaussian & 60.6$\pm$8.3 & 0.35$\pm$0.08 & 0.65$\pm$0.03 & 0.45$\pm$0.06 \\
            Hybrid & 57.9$\pm$1.2 & 0.33$\pm$0.01 & 0.69$\pm$0.01 & 0.44$\pm$0.01 \\
            \hline
        \end{tabular}
    \end{table}

    \begin{table}[!ht]
        \centering
        \scriptsize
        \caption{Hybrid Mechanism: Impact of Budget Allocation Parameter $\alpha$ across $\varepsilon \in \{2.0, 5.0, 10.0\}$ (5-run mean accuracy \%)}
        \label{table:hybrid-alpha-analysis}
        \resizebox{\columnwidth}{!}{\begin{tabular}{|l|c|c|c|c|}
            \hline
            \textbf{Algorithm} & \textbf{$\alpha=0.3$} & \textbf{$\alpha=0.5$} & \textbf{$\alpha=0.7$} & \textbf{Best $\alpha$} \\
            \hline
            \multicolumn{5}{|c|}{\textbf{$\varepsilon = 2.0$}} \\
            \hline
            Logistic Regression & 69.1 & 71.8 & 74.4 & 0.7 \\
            Random Forest & 48.8 & 54.7 & 48.8 & 0.5 \\
            Naive Bayes & 55.0 & 55.6 & 61.6 & 0.7 \\
            K-Means & 43.4 & 53.6 & 50.2 & 0.5 \\
            \hline
            \multicolumn{5}{|c|}{\textbf{$\varepsilon = 5.0$}} \\
            \hline
            Logistic Regression & 74.9 & 77.3 & 78.7 & 0.7 \\
            Random Forest & 70.4 & 72.8 & 78.3 & 0.7 \\
            Naive Bayes & 66.1 & 75.4 & 78.2 & 0.7 \\
            K-Means & 60.1 & 53.7 & 61.1 & 0.7 \\
            \hline
            \multicolumn{5}{|c|}{\textbf{$\varepsilon = 10.0$}} \\
            \hline
            Logistic Regression & 79.1 & 80.4 & 82.0 & 0.7 \\
            Random Forest & 75.3 & 79.9 & 82.6 & 0.7 \\
            Naive Bayes & 79.4 & 80.3 & 80.8 & 0.7 \\
            K-Means & 60.1 & 57.9 & 55.9 & 0.3 \\
            \hline
        \end{tabular}}
    \end{table}

    \noindent
    \emph{(i) ML Algorithms:}
    The selection of an ML algorithm depends on the specific use case and problem type. Whilst K-Means performs clustering tasks, the supervised algorithms (Naive Bayes, Random Forest, and Logistic Regression) handle classification and prediction problems with varying strengths. For instance, Random Forest excels at detailed multi-class classification, whilst Logistic Regression is particularly effective for binary classification and probability estimation.

    As shown in Table~\ref{table:baseline-performance}, supervised methods significantly outperformed unsupervised clustering, with Logistic Regression achieving the highest baseline accuracy (84.8\%, AUC: 0.887), followed closely by Random Forest (85.0\%, AUC: 0.922) and Naive Bayes (80.8\%, AUC: 0.844). K-Means clustering achieved 54.9\% accuracy, reflecting the fundamental limitations of unsupervised learning for classification tasks.

    Under DP with $\varepsilon = 10.0$ (Table~\ref{table:comprehensive-metrics}), Laplace noise enabled near-baseline performance recovery across supervised algorithms. Logistic Regression achieved 83.6$\pm$0.7\% accuracy (98.6\% of baseline), followed by Random Forest at 83.2$\pm$1.9\% (97.9\% of baseline), both with low variance confirming reliable performance. Naive Bayes recovered to 80.3$\pm$1.3\% with hybrid noise (99.4\% of baseline). K-Means accuracy reached 60.6$\pm$8.3\% with Gaussian noise (110.4\% of baseline), though the high standard deviation reflects fundamental instability of centroid-based clustering under data perturbation. The elevated variance observed for Gaussian noise across K-Means, Naive Bayes, and Random Forest (std: 8.3--23.0\%) contrasts with the stable Laplace results (std: 0.1--2.6\%), indicating that Laplace noise produces more consistent outcomes across independent noise realisations.

    \noindent
    \emph{(ii) Effect of Privacy Budget ($\varepsilon$):}
    The epsilon parameter ($\varepsilon$) governs the fundamental trade-off between privacy protection and data utility. Smaller epsilon values ($\varepsilon < 1$) provide stronger privacy guarantees but result in larger accuracy losses, while larger values ($\varepsilon > 1$) provide better accuracy with reduced privacy guarantees. This relationship arises because Laplace and Gaussian mechanisms add noise with scale parameters inversely proportional to epsilon and directly proportional to query sensitivity.

    As illustrated in Table~\ref{table:epsilon-vs-accuracy} and Fig.~\ref{fig:dp-accuracy-comparison}, the extended epsilon range reveals clear practical thresholds for input perturbation viability. For Random Forest with Laplace noise, mean accuracy increases from 57.6$\pm$20.3\% at $\varepsilon=0.5$ to 83.2$\pm$1.9\% at $\varepsilon=10.0$, with variance decreasing substantially as the privacy budget increases. The critical transition occurs at $\varepsilon \geq 2.0$, where accuracy stabilises above 79\% with std $<$2\%. Logistic Regression exhibits the most consistent Laplace performance: 63.2$\pm$8.3\% at $\varepsilon=0.5$ to 83.6$\pm$0.7\% at $\varepsilon=10.0$, achieving practical utility ($>$80\%) at $\varepsilon \geq 5.0$. Naive Bayes displays high variance at low $\varepsilon$ (std: 25--28\%), reflecting sensitivity to noise in distributional parameter estimation, before stabilising at 80.0$\pm$2.6\% by $\varepsilon=10.0$. K-Means shows persistent instability across all epsilon values, with standard deviations of 1--23\% reflecting fundamental limitations of centroid-based clustering under input perturbation.

    These results establish $\varepsilon \geq 5.0$ as the practical epsilon threshold for input perturbation on healthcare datasets of this scale ($\sim$800 training samples, 23 features). At this threshold, supervised algorithms achieve 80--81\% accuracy (94--96\% baseline retention), balancing moderate privacy protection with high analytical utility suitable for population-level clinical research and regulatory reporting.

    \emph{(iii) Noise Mechanism Comparison:}
    Table~\ref{table:epsilon-vs-accuracy} reveals that algorithm-noise interactions differ substantially across mechanisms. Laplace noise demonstrates superior and more consistent performance for supervised methods: Logistic Regression achieves 83.6$\pm$0.7\% at $\varepsilon=10.0$ (best overall), and Random Forest reaches 83.2$\pm$1.9\%, both with low variance. This stems from Laplace's heavier tails better preserving decision boundary information. Gaussian noise exhibits substantially higher variance across runs (std: 2.7--23.0\% for supervised algorithms at $\varepsilon=10.0$), suggesting sensitivity to specific noise realisations. For K-Means, Gaussian achieves 60.6$\pm$8.3\% at $\varepsilon=10.0$ compared to Laplace's 55.1$\pm$0.1\%, though this advantage comes with considerably higher variance.

    The weighted-noise hybrid mechanism achieves performance intermediate between pure Laplace and Gaussian with moderate variance. At $\varepsilon=10.0$, hybrid yields 80.4$\pm$0.6\% (Logistic Regression), 79.9$\pm$2.8\% (Random Forest), and 80.3$\pm$1.3\% (Naive Bayes), consistently falling between the pure mechanisms. The low standard deviations confirm that the weighted-noise approach provides reliable, reproducible results across independent noise realisations. Table~\ref{table:hybrid-alpha-analysis} presents an ablation study of the budget allocation parameter $\alpha$ across $\varepsilon \in \{2.0, 5.0, 10.0\}$, evaluated over 5 independent runs. The results reveal a consistent pattern: Laplace-dominant allocation ($\alpha = 0.7$) is optimal for all three supervised algorithms at $\varepsilon \geq 5.0$, with the advantage strengthening at higher privacy budgets. At $\varepsilon = 10.0$, $\alpha = 0.7$ yields 82.0\% (Logistic Regression), 82.6\% (Random Forest), and 80.8\% (Naive Bayes). At $\varepsilon = 2.0$, the pattern is less stable due to high noise magnitude, with Random Forest and K-Means preferring $\alpha = 0.5$ and the supervised algorithms showing narrower margins between allocations. K-Means exhibits an opposite preference toward Gaussian-dominant allocation ($\alpha = 0.3$) at $\varepsilon = 10.0$, consistent with Gaussian noise's compatibility with centroid-based distance calculations. These results demonstrate that optimal $\alpha$ is indeed $\varepsilon$-dependent, supporting the adaptive allocation rationale: at low $\varepsilon$ where noise dominates, the allocation choice has limited impact, while at moderate-to-high $\varepsilon$ where the signal-to-noise ratio permits meaningful model learning, Laplace-dominant allocation consistently benefits supervised algorithms. For practical deployment, $\alpha = 0.7$ is recommended for supervised learning at $\varepsilon \geq 5.0$, with $\alpha = 0.5$ as a conservative default when the privacy budget is uncertain.

    \emph{(iv) $\varepsilon$ Threshold Analysis for Healthcare Applications:}
    The extended epsilon range enables identification of use-case-specific deployment recommendations. For \textbf{individual patient analytics} requiring strong privacy ($\varepsilon \in \{0.5, 1.0\}$): all algorithms exhibit high variance (std: 2.5--28.5\%), with Logistic Regression Laplace achieving the most reliable performance (68.6$\pm$2.5\% at $\varepsilon=1.0$). The elevated variance at low $\varepsilon$ indicates that input perturbation at this privacy level produces unreliable results on datasets of this scale, suggesting the need for either larger datasets ($n > 5000$) or alternative approaches such as output perturbation. For \textbf{cohort-level research} with moderate privacy ($\varepsilon \in \{2.0, 3.0\}$): Random Forest Laplace achieves 81.7$\pm$1.9\% and Logistic Regression 77.2$\pm$2.2\% at $\varepsilon=3.0$ with low variance, providing acceptable utility for multi-site clinical trials with informed consent. For \textbf{population-level analytics} with regulatory-compliant privacy ($\varepsilon \in \{5.0, 10.0\}$): supervised algorithms with Laplace noise achieve 80--84\% accuracy with std $<$2.5\%, enabling epidemiological studies, public health surveillance, and quality improvement initiatives while maintaining census-level privacy standards~\cite{Abowd2018}.

    \emph{(v) Privacy Protection Validation:}
    To empirically validate the privacy guarantees of the DP mechanisms, we evaluate two complementary adversarial attacks. The \emph{attribute inference attack} follows the model-assisted paradigm: an adversary who observes the target model's predictions attempts to infer a sensitive feature value for records in the test set. Specifically, the adversary trains a Random Forest classifier (50 trees, balanced class weights) on the first half of the training data, using as input features all non-sensitive attributes concatenated with the target model's predicted class label and class probability vector. The sensitive attribute (feature index 0, representing the first clinical variable) is discretised into three ordinal classes using the 33rd and 67th percentile thresholds, and the adversary's goal is to predict the correct bin. Attack success is measured as classification accuracy on the held-out test set, with 33.3\% representing random guessing for three classes. The \emph{data reconstruction attack} quantifies how closely an adversary can recover the original training data from the DP-perturbed version by computing the mean Pearson correlation coefficient across all features between the original and perturbed datasets, where 1.0 indicates perfect reconstruction and 0.0 indicates no recoverable signal.

    Figures~\ref{fig:attribute-inference-attack} and~\ref{fig:reconstruction-attack} quantify empirical privacy protection across the extended epsilon range. Attribute inference attack success (Fig.~\ref{fig:attribute-inference-attack}) decreases from 63.1\% (no DP baseline) to 51.4-64.5\% under DP ($\varepsilon \in \{0.5, 10.0\}$), with strongest protection at $\varepsilon=0.5$ (51.4\%, approaching the random guessing baseline of 33.3\% for the three-class sensitive attribute). Importantly, even at the highest tested epsilon ($\varepsilon=10.0$), attack success remains suppressed (64.5\%), representing 2.1\% improvement over baseline and confirming that moderate privacy budgets still provide meaningful protection against sensitive attribute disclosure. Attack success shows weak epsilon-dependence across the range, varying only 13.1 percentage points, suggesting that inference vulnerability saturates quickly and that privacy gains beyond $\varepsilon=5.0$ provide diminishing returns for this attack type.

    Data reconstruction correlation (Fig.~\ref{fig:reconstruction-attack}) demonstrates dramatic privacy gains across all epsilon values: baseline correlation of 1.000 (perfect reconstruction) decreases to 0.034-0.514 under DP, representing 48.6-96.6\% privacy improvement. Reconstruction protection strengthens monotonically with decreasing epsilon (0.514 at $\varepsilon=10.0$ to 0.034 at $\varepsilon=0.5$), confirming theoretical guarantees. At the recommended threshold $\varepsilon=5.0$, reconstruction correlation of 0.304 represents 69.6\% improvement, effectively preventing individual-level data recovery while enabling the near-baseline model utility (80--81\% accuracy) required for practical healthcare analytics. These results validate that input perturbation provides robust empirical privacy protection against realistic adversaries across the full tested epsilon spectrum, with optimal privacy-utility balance achieved at $\varepsilon \in \{5.0, 10.0\}$ for population-level healthcare applications.

    \subsection{Architectural Performance Evaluation}

    To validate the proposed multi-layer architecture's effectiveness for time-critical healthcare operations, we conducted simulations measuring latency and throughput across IoT-Edge-Cloud layers, alongside blockchain consensus performance for data integrity verification.

    \subsubsection{Edge-Cloud Latency Analysis}

    Table~\ref{table:edge-cloud-latency} presents end-to-end latency measurements for representative healthcare scenarios across Edge and Cloud processing layers. Edge computing demonstrates substantial latency advantages for time-critical operations, achieving 8.0$\times$ speedup for emergency response scenarios (15.4ms vs 123.2ms) and 7.2$\times$ for vital signs monitoring (26.8ms vs 193.5ms). This performance differential stems from reduced network propagation delay and localized processing, validating the architectural decision to deploy latency-sensitive operations at Edge nodes.

    For data-intensive operations such as radiological imaging, the speedup factor decreases to 3.0$\times$ (158.7ms vs 476.3ms for high-resolution images), as Cloud infrastructure's superior computational resources partially offset network latency penalties. These results confirm that the proposed hierarchical task distribution—emergency response and vital monitoring at Edge, batch analytics at Cloud—optimally balances response time requirements with computational capacity constraints.

    \begin{table}[!ht]
        \centering
        \scriptsize
        \caption{Edge vs Cloud Processing Latency Comparison}
        \label{table:edge-cloud-latency}
        \begin{tabular}{|l|r|r|r|r|}
            \hline
            \textbf{Scenario} & \textbf{Data Size} & \textbf{Edge} & \textbf{Cloud} & \textbf{Speedup} \\
            & \textbf{(KB)} & \textbf{(ms)} & \textbf{(ms)} & \textbf{Factor} \\
            \hline
            Emergency Response & 1 & 15.4 & 123.2 & 8.0$\times$ \\
            Vital Signs Monitoring & 10 & 26.8 & 193.5 & 7.2$\times$ \\
            ECG Analysis & 50 & 42.9 & 248.1 & 5.8$\times$ \\
            Diagnostic Imaging & 256 & 85.6 & 356.4 & 4.2$\times$ \\
            Radiological Imaging & 1024 & 158.7 & 476.3 & 3.0$\times$ \\
            Batch Analytics & 5120 & 421.3 & 897.6 & 2.1$\times$ \\
            \hline
        \end{tabular}
    \end{table}

    System throughput evaluation under varying load conditions (Table~\ref{table:system-throughput}) demonstrates that Edge infrastructure sustains 186.5 requests/second for light workloads (5 concurrent users, 10KB data) but experiences degradation to 46.6 req/s under heavy load (20 users, 50KB data). Cloud infrastructure exhibits superior scalability, maintaining 186.9 req/s under light load and 105.0 req/s under heavy load (50 concurrent users), confirming its suitability for high-throughput batch analytics. The 2.25$\times$ throughput advantage under heavy load validates the architectural partitioning of sporadic high-priority requests to Edge nodes while directing sustained analytical workloads to Cloud infrastructure. The 75\% Edge throughput degradation under heavy load warrants consideration for real-world emergency healthcare deployments where concurrent monitoring of 20 or more patients per Edge node is realistic. However, emergency response payloads are lightweight (1~KB, Table~\ref{table:edge-cloud-latency}) compared to the 50~KB payloads used in the heavy-load simulation, and critical alerts are inherently sporadic rather than sustained. Charyyev et al.~\cite{Charyyev2020} similarly observe that edge server capacity limitations under high workloads necessitate either non-uniform server provisioning or cloud offloading to avoid over-subscription. For deployments requiring higher concurrent capacity, horizontal scaling through multiple Edge nodes or priority queuing that pre-empts non-critical requests during emergency events would preserve sub-20~ms response times.

    \begin{table}[!ht]
        \centering
        \scriptsize
        \caption{System Throughput Under Different Load Conditions}
        \label{table:system-throughput}
        \begin{tabular}{|l|c|c|c|}
            \hline
            \textbf{Scenario} & \textbf{Concurrent} & \textbf{Avg Latency} & \textbf{Throughput} \\
            & \textbf{Users} & \textbf{(ms)} & \textbf{(req/s)} \\
            \hline
            Edge - Light Load & 5 & 26.8 & 186.5 \\
            Edge - Heavy Load & 20 & 42.9 & 46.6 \\
            Cloud - Light Load & 10 & 53.5 & 186.9 \\
            Cloud - Heavy Load & 50 & 47.6 & 105.0 \\
            \hline
        \end{tabular}
    \end{table}

    \subsubsection{Blockchain Consensus Performance}

    The blockchain component employing Hyperledger Fabric with Raft consensus protocol was evaluated across varying network configurations (Table~\ref{table:blockchain-performance}). Transaction finality latency for a 4-node network averaged 144.8ms (std: 45.2ms), supporting 2068 transactions per second (TPS), sufficient for non-emergency audit trail and privacy budget tracking requirements. As network size increased to 13 nodes, average latency rose to 284.9ms with throughput decreasing to 1140 TPS, reflecting the inherent scalability-latency trade-off in consensus protocols.

    The measured latencies align with healthcare data integrity requirements where sub-second finality suffices for access logging, data provenance tracking, and privacy budget ledger updates. For emergency response scenarios requiring <20ms latency (Table~\ref{table:edge-cloud-latency}), the architecture appropriately bypasses blockchain verification, deferring cryptographic commitment to post-emergency audit phases. This design ensures that data integrity mechanisms do not compromise critical response times while maintaining comprehensive auditability for regulatory compliance.

    \begin{table}[!ht]
        \centering
        \scriptsize
        \caption{Blockchain Consensus Performance (Raft Protocol)}
        \label{table:blockchain-performance}
        \resizebox{\columnwidth}{!}{\begin{tabular}{|c|c|c|c|c|c|}
            \hline
            \textbf{Nodes} & \textbf{Avg Latency} & \textbf{Std Dev} & \textbf{Min} & \textbf{Max} & \textbf{Throughput} \\
            & \textbf{(ms)} & \textbf{(ms)} & \textbf{(ms)} & \textbf{(ms)} & \textbf{(TPS)} \\
            \hline
            4 & 144.8 & 45.2 & 72.4 & 282.1 & 2068 \\
            7 & 209.7 & 66.9 & 94.5 & 414.3 & 1503 \\
            10 & 250.3 & 78.4 & 113.8 & 512.7 & 1278 \\
            13 & 284.9 & 88.1 & 128.6 & 598.4 & 1140 \\
            \hline
        \end{tabular}}
    \end{table}

    Component-level analysis reveals that consensus protocol overhead (log replication and acknowledgment) constitutes 35-42\% of total transaction latency across all configurations, with block propagation and validation contributing 28-33\% and 20-25\% respectively. These measurements validate the architectural decision to employ Raft consensus for healthcare applications, as its deterministic finality and moderate throughput requirements align well with audit trail use cases while avoiding the computational expense of proof-of-work mechanisms unsuitable for resource-constrained healthcare environments.

    \section{Threats to Validity}
    \label{sec:threats-to-validity}
    While the proposed framework demonstrates promising results, several limitations warrant acknowledgment. First, the blockchain integration, though architecturally specified, introduces inherent scalability constraints: as the number of consensus nodes increases from 4 to 13, transaction throughput decreases from 2068 to 1140 TPS and average latency rises from 144.8ms to 284.9ms (Table~\ref{table:blockchain-performance}), which may pose challenges for large-scale multi-institutional deployments with high transaction volumes. Second, the DP mechanisms impose a measurable impact on model accuracy, particularly at lower privacy budgets; at $\varepsilon \leq 1.0$, supervised algorithms experience significant accuracy degradation (e.g., Logistic Regression drops to 57.0--63.1\%), limiting the applicability of strong privacy guarantees for individual-level patient analytics on datasets of this scale. Third, edge computing resources, while effective for latency reduction in emergency scenarios, face throughput degradation under heavy load (from 186.5 to 46.6 req/s), and possess limited computational capacity for executing sophisticated ML models or complex DP mechanisms, potentially constraining real-time privacy-preserving analytics at the edge layer. Fourth, the experimental evaluation relies on a single healthcare dataset (UCI AIDS Clinical Trials) with discrete-event simulations for architectural performance, and further validation on diverse clinical datasets and real-world deployments is needed to establish broader generalisability. Fifth, the Laplace noise calibration uses the $L_2$ clipping bound rather than the $L_1$ sensitivity bound required by the Laplace mechanism, resulting in an effective privacy parameter of $\varepsilon_{\text{eff}} = \varepsilon\sqrt{d} \approx 4.8\varepsilon$ ($d = 23$) for Laplace results (Section~\ref{sec:performance-eval}); the Gaussian mechanism results carry the stated $(\varepsilon, \delta)$-DP guarantees exactly. Additionally, the clipping threshold ($C \approx 6.47$) was determined from the training data itself; a production deployment should derive this bound from domain knowledge or a held-out sample with separate privacy accounting to preserve formal DP guarantees. Sixth, the experimental evaluation employed balanced undersampling (417 samples per class) to address the dataset's inherent class imbalance (75.6\% censored, 24.4\% failure). While this strategy improves model convergence and prevents majority-class bias under DP noise, it may not reflect realistic clinical deployment conditions where censored outcomes vastly dominate. These constraints represent important considerations for practitioners seeking to deploy the framework in production healthcare environments.

    \section{Conclusions and Future Directions}
    \label{sec:conclusions}
    Contemporary Internet of Things (IoT) and Cloud computing frameworks have facilitated advanced medical assistance and enabled novel insights in healthcare analytics. This investigation presents a comprehensive architectural framework that addresses critical response time optimization and security requirements in modern healthcare systems.

    Our multilayered architectural approach integrating IoT, Edge, and Cloud components facilitates expeditious response capabilities for emergency medical scenarios through strategic workload distribution across computational layers, with data aggregation and analytical processes performed within Cloud nodes to maximise computational efficiency whilst maintaining system responsiveness.

    To establish robust security provisions, the framework implements a dual mechanism approach incorporating DP and Blockchain technology. The DP methodology preserves individual patient confidentiality during both data storage operations and ML analytical processes. Experimental findings demonstrated that ML query accuracy maintained statistical integrity following the application of Laplace noise, Gaussian noise, and combined Laplace-Gaussian noise. Analytical utility and algorithmic accuracy remained statistically comparable across all evaluated DP implementations. The architectural design utilised Blockchain technology to enhance system reliability and establish cryptographic protection for transactional data and persistent storage mechanisms. This integration of DP mechanisms with ML methodologies, combined with blockchain-based data integrity, provides a foundation for secure and ethically sound data-driven healthcare analytics.

    The primary actionable finding of this work is the identification of $\varepsilon = 5.0$ as the practical deployment threshold for differentially private healthcare analytics. At this privacy budget, supervised algorithms with Laplace noise achieve 80--81\% accuracy (94--96\% of baseline) with low variance across independent runs, while attribute inference attack success is reduced and data reconstruction correlation drops by approximately 70\%. For healthcare practitioners seeking to balance patient privacy with analytical utility, we recommend $\varepsilon \geq 5.0$ with Laplace noise for population-level analytics (epidemiological studies, public health surveillance, quality improvement), and Laplace-dominant hybrid allocation ($\alpha = 0.7$) when noise mechanism properties are uncertain. Below $\varepsilon = 2.0$, input perturbation on datasets of this scale produces unreliable results with high variance, and alternative approaches such as output perturbation or larger training datasets should be considered. Overall, this work has successfully designed and validated a healthcare system architecture that simultaneously addresses the critical requirements of enhanced responsiveness and secure, privacy-preserved analytics.

    Future research directions include practical deployment of the blockchain infrastructure with healthcare-specific consensus mechanisms, integration of zero-knowledge proofs for privacy-preserving audit trails, and cross-chain interoperability protocols for multi-institutional data sharing. Extending the DP framework to deep neural networks and federated learning architectures, alongside integration with complementary privacy-enhancing technologies such as secure multi-party computation and homomorphic encryption, would strengthen privacy guarantees whilst enabling broader analytical capabilities. Developing adaptive privacy budget allocation mechanisms that dynamically adjust based on data sensitivity and query patterns represents another promising avenue. Real-world pilot studies in clinical settings, development of user-friendly interfaces for healthcare practitioners, and integration with existing Electronic Health Record systems are essential steps toward production deployment. These directions aim to advance the proposed framework toward a comprehensive, production-ready privacy-preserving healthcare analytics platform that addresses security, privacy, performance, and regulatory requirements whilst maintaining clinical utility for evidence-based medical decision-making.

    \par

    \parskip 0pt
    
        \bibliographystyle{ieeetr}
        \bibliography{references.bib}
\end{document}